\documentclass[twocolumn, trackchanges]{aastex701}

\DeclareUnicodeCharacter{2212}{-}

\begin{document}

\title{The SOFIA Massive (SOMA) Radio Survey. II. Radio Emission from High-Luminosity Protostars}

\author[orcid=0000-0001-8169-1437, gname='Francisco', sname='Sequeira-Murillo']{Francisco Sequeira-Murillo}
\affiliation{Department of Astronomy, University of Wisconsin-Madison, 475 N. Charter St., Madison, WI 53703, USA}
\email[show]{sequeiramuri@wisc.edu}

\author[orcid=0000-0001-8596-1756, gname='Viviana', sname='Rosero']{Viviana Rosero}
\affiliation{Cahill Center for Astronomy and Astrophysics, MC 249-17, California Institute of Technology, Pasadena, CA 91125, USA}
\affiliation{National Radio Astronomy Observatory, 1003 Lópezville Rd., Socorro, NM 87801, USA} 
\affiliation{Space Science Institute, 4750 Walnut Street, Suite 205, Boulder, CO 80301, USA}
\email[show]{vrosero@caltech.edu}   

\author[orcid=0000-0003-1111-8066, gname='Joshua', sname='Marvil']{Joshua Marvil}
\affiliation{National Radio Astronomy Observatory, 1003 Lópezville Rd., Socorro, NM 87801, USA}
\email{jmarvil@nrao.edu}  

\author[orcid=0000-0002-3389-9142, gname='Jonathan C.', sname='Tan']{Jonathan C. Tan}
\affiliation{Department of Space, Earth \& Environment, Chalmers University of Technology, 412 93 Gothenburg, Sweden}
\affiliation{Department of Astronomy, University of Virginia, Charlottesville, Virginia 22904, USA} 
\email{jonathan.tan@chalmers.se}  

\author[orcid=0000-0003-4040-4934, gname='Ruben', sname='Fedriani']{Ruben Fedriani}
\affiliation{Instituto de Astrofísica de Andalucía, CSIC, Glorieta de la Astronomía s/n, 18008 Granada, Spain}
\affiliation{Department of Space, Earth \& Environment, Chalmers University of Technology, 412 93 Gothenburg, Sweden} 
\email{fedriani@iaa.es}  

\author[orcid=0000-0001-7511-0034, gname='Yichen', sname='Zhang']{Yichen Zhang}
\affiliation{State Key Laboratory of Dark Matter Physics, School of Physics and Astronomy, Shanghai Jiao Tong University, Shanghai 200240, People’s Republic of China}
\affiliation{Department of Astronomy, University of Virginia, Charlottesville, Virginia 22904, USA}
\email{yichen.zhang@sjtu.edu.cn}

\author[orcid=0009-0007-4080-9807, gname='Azia', sname='Robinson']{Azia Robinson}
\affiliation{Department of Physics and Astronomy, Agnes Scott College, 141 E. College Ave. Decatur, GA, 30030}
\email{a.k.ro2548@gmail.com}

\author[orcid=0000-0003-1602-6849, gname='Prasanta', sname='Gorai']{Prasanta Gorai}
\affiliation{Rosseland Centre for Solar Physics, University of Oslo, PO Box 1029 Blindern, 0315, Oslo, Norway}
\affiliation{Institute of Theoretical Astrophysics, University of Oslo, PO Box 1029 Blindern, 0315, Oslo, Norway}
\email{prasanta.gorai@astro.uio.no}

\author[orcid=0000-0002-6907-0926, gname='Kei E. I.', sname='Tanaka']{Kei E. I. Tanaka}
\affiliation{Department of Earth and Planetary Sciences, Institute of Science Tokyo, Meguro, Tokyo 152-8551, Japan}
\email{kei.tanaka@eps.sci.isct.ac.jp}

\author[orcid=0000-0001-7378-4430, gname='James M.', sname='De Buizer']{James M. De Buizer}
\affiliation{Carl Sagan Center for Research, SETI Institute, Mountain View, CA, USA}
\email{jdebuizer@seti.org}

\author[orcid=0000-0003-3315-5626, gname='Maria T.', sname='Beltrán']{Maria T. Beltrán}
\affiliation{INAF-Osservatorio Astrofisico di Arcetri, Largo E. Fermi 5, I-50125 Firenze, Italy}
\email{maria.beltran@inaf.it}

\author[orcid=0000-0001-9857-1853, gname='Ryan D.', sname='Boyden']{Ryan D. Boyden}
\affiliation{Department of Astronomy, University of Virginia, Charlottesville, Virginia 22904, USA}
\email{xhs4wj@virginia.edu}

\defcitealias{Bruizer_2017}{SOMA I}
\defcitealias{Liu_2019}{SOMA II}
\defcitealias{Liu_2020}{SOMA III}
\defcitealias{Fedriani_2023}{SOMA IV}
\defcitealias{Telkamp_2025}{SOMA V}
\defcitealias{Rosero_2019}{SOMA Radio I}
\defcitealias{Zhang_2018}{ZT18}
\defcitealias{Tanaka_2016}{TTZ16}

\begin{abstract}

We present centimeter continuum observations of seven high luminosity massive protostars and their surrounding sources in regions with multiple targets, as part of the SOFIA Massive (SOMA) Star Formation Survey.
With data from the Very Large Array and the Australia Telescope Compact Array, we analyze the spectral index, morphology and multiplicity of the detected radio sources. The high-sensitivity, high-resolution observations allow us to resolve many sources; 65$\%$ of the reported sources are resolved at least within the synthesized beam. 
We report fifteen new detections, thirteen of which are entirely new and two that have counterparts at other wavelengths but are detected here for the first time at radio frequencies.
We use the observations to build radio spectral energy distributions (SEDs) to calculate spectral indices. With radio morphologies and the spectral indices, we give assessments on the nature of the sources, highlighting six sources that display a radio jet-like morphology and a spectral index consistent with ionized jets. Combining with the SOMA Radio I sample, we present the radio - bolometric luminosity relation, especially probing the regime from $L_{\rm bol}\sim 10^4$ to $10^6\:L_\odot$. Here we find a steep rise in radio luminosity, which is expected by models that transition from shock ionization to photoionization.

\end{abstract}


\section{Introduction} \label{sec:intro}

Massive stars impact many processes in astrophysics, yet crucial aspects of their formation remain uncertain due to observational difficulties given by their relative rarity and large distances (typically $>1\:$kpc), crowded environments and high extinctions \citep{2014prpl.conf..149T}. Various theories have been proposed to explain the formation of massive stars, but there is no general consensus on the basic nature of the formation mechanism. Theories range from the Turbulent Core Accretion (TCA) model \citep{2003ApJ...585..850M} to Competitive Accretion \citep[e.g.,][]{2001MNRAS.323..785B,2010ApJ...709...27W,2022MNRAS.512..216G} to Protostellar Collisions \citep{1998MNRAS.298...93B}.

High-sensitivity centimeter continuum observations can provide valuable understanding of the earliest and most obscured phases of high-mass star formation. \cite{Tanaka_2016} (hereafter \citetalias{Tanaka_2016}) developed models to predict the ionization structures and centimeter continuum emission properties using initial parameters from a physical model based on TCA, including infall envelope, disk and outflow properties and their predicted thermal spectral energy distribution (SED) \citep{2011ApJ...733...55Z,2013ApJ...766...86Z,2014ApJ...788..166Z,Zhang_2018}. \cite{2016ApJS..227...25R,2019ApJ...880...99R} demonstrated that many centimeter continuum sources have morphologies and parameters indicative of ionized jets, which is consistent with the results from the free-free emission models developed by \citetalias{Tanaka_2016}. This makes radio continuum emission highly relevant for constraining the ionizing luminosity of a protostar and helping to develop and test theoretical models for the evolutionary sequence of massive star formation.

In order to carry out such tests one needs a sample of massive protostars that ideally spans a range of core properties, evolutionary stages and environments. The SOFIA Massive (SOMA) Star Formation Survey (PI: J. Tan) was developed with such goals in mind. It aims to characterize a sample of $\geq$ 50 high and intermediate-mass protostars over a range of initial core masses, evolutionary stages and environments using SOFIA-FORCAST $\sim$ 10–40 $\mu$m data. In Paper I of the survey \citep{Bruizer_2017} (hereafter \citetalias{Bruizer_2017}), the first eight sources were presented, which were mostly massive protostars. In Paper II \citep{Liu_2019} (hereafter \citetalias{Liu_2019}), seven high-luminosity sources were presented, representing some of the most massive protostars in the survey. In Paper III \citep{Liu_2020} (hereafter \citetalias{Liu_2020}), 14 intermediate-mass sources were presented. In Paper IV \citep{Fedriani_2023} (hereafter \citetalias{Fedriani_2023}), 11 isolated sources, based on the 37 $\mu$m imaging, were presented. In Paper V, \cite{Telkamp_2025} (hereafter \citetalias{Telkamp_2025}) seven regions of relatively clustered massive star formation were studied.

The SOMA Radio Survey was developed with the goal of obtaining cm radio continuum observations of the SOMA sample, i.e., to have more comprehensive wavelength coverage of ``extended spectral energy distributions'' (E-SEDs) that constrains ionized gas properties and thus help break degeneracies that arise in infrared-only SED inferred properties from the standard SOMA analysis based on the \cite{Zhang_2018} (hereafter \citetalias{Zhang_2018}) TCA model grid. In SOMA Radio Paper I \citep{Rosero_2019} (hereafter \citetalias{Rosero_2019}), the first eight sources of \citetalias{Bruizer_2017} were observed with the Karl G. Jansky Very Large Array (VLA) and the centimeter continuum data presented and analyzed, including comparison to the \citetalias{Tanaka_2016} models for free-free emission.

In this paper, second in the radio series, we present ATCA and VLA data of the seven high-luminosity regions of \citetalias{Liu_2019}, containing a total of nine sources, i.e., after adding IRAS 16562-3959 N and G305.20+0.21 A that were reported and analyzed in \citetalias{Liu_2020}.

Our approach follows the general methods developed in \citetalias{Rosero_2019} to analyze the data. In particular, we construct radio SEDs by reducing the centimeter continuum observations, measuring the fluxes from the images, and analyzing the morphology and multiplicity of each region. We define multiplicity in a region when it  contains two or more 5$\sigma$ detections that are not necessarily associated with each other. We note that due to improvements in the infrared SED analysis, including a new algorithm to calculate the optimal aperture size in an unbiased and reproducible way for extended sources, updates in the SED fitting tool and revised methods assessing uncertainties in background subtracted fluxes, we use the infrared inferred results (e.g., bolometric luminosities) from \citetalias{Telkamp_2025}, rather than the results from \citetalias{Liu_2019}.

The paper is organized as follows: methodology and information regarding the observations are presented in \S\ref{sec:methods}. The observational results for each source are presented in \S\ref{subsec:results}, while the analysis and discussion of the sample are presented in \S\ref{analysis}. A summary and our conclusions are presented in \S\ref{summary}.

\section{Methods}\label{sec:methods}

The SOMA Star Formation Survey sample is defined by SOFIA-FORCAST observations (i.e., from $\sim$7 to $40\:\mu$m). The sample we analyze in this paper are part of the protostars presented by \citetalias{Liu_2019}, as well as sources IRAS 16562-3959 N and G305.20+0.21A that were reported in \citetalias{Liu_2020}, as an addendum to the sources in the \citetalias{Liu_2019} regions. Thus a total of nine protostars in seven target regions will be analyzed: G45.12+0.13, G309.92+0.48, G35.58-0.03, IRAS 16562-3959, IRAS 16562-3959 N, G305.20+0.21, G305.20+0.21 A, G49.27-0.34 and G339.88-1.26. The radio observations presented here are our own VLA observations for regions G45.12+0.13, G35.58-0.03 and G49.27-0.34. For G305.20+0.21 and G309.92+0.48 we present our own ATCA observations, and for regions IRAS 16562-3959 and G339.88-1.26, we used ATCA observations reported by \cite{2016ApJ...826..208G} and \cite{2016MNRAS.460.1039P}, respectively.

The data analyzed in this work are summarized in Table \ref{tab:SOMA_Sources}. Column 1 gives region name; columns 2, 3, and 4 provide band frequency, R.A., and Decl., respectively; columns 5 and 6 give synthesized beam size and position angle (PA), and the rms of the resulting images. The distance, as adopted by \citetalias{Liu_2019}, as well as the bolometric luminosities and the isotropic bolometric luminosity evaluated by \citetalias{Telkamp_2025}, are shown in columns 7, 8 and 9, respectively. A list of phase calibrators used in the VLA observations at 1.3 and 6 cm is given in Table \ref{tab:VLA_Calibrators}.

\startlongtable
\begin{deluxetable*}{cccccccccc}
    \tabletypesize{\scriptsize}
    \tablewidth{0pt}
    \tablecaption{SOMA Sources: Radio Continuum Data. \label{tab:SOMA_Sources}} 
    \tablehead{\colhead{Region} & \colhead{Frequency Band} & \colhead{R.A} & \colhead{Decl.} & \colhead{Beam Size} & \colhead{rms} & \colhead{$d$\small\textsuperscript{c}} & \colhead{$L_{\rm bol, iso}$\small\textsuperscript{d}}  & \colhead{$L_{\rm bol}$\small\textsuperscript{d}}\\
    \colhead{} & \colhead{(GHz)} & \colhead{(J2000)} & \colhead{(J2000)} & \colhead{($^{\prime\prime}\times ^{\prime\prime}$, degree)} & \colhead{($\mu$Jy beam$^{-1}$)} & \colhead{(kpc)} & \colhead{($L_{\odot}$)} & \colhead{($L_{\odot}$)}}
    \startdata
    G45.12+0.13\small\textsuperscript{a} & 4.0 $-$ 8.0 & 19 13 27.86 & +10 53 36.6  & 0.26 $\times$ 0.25, $-$45.5 & 35 & 7.40 & $4.4^{+4.4}_{-2.2} \times 10^{5}$ & $8.0^{+6.5}_{-3.6} \times 10^{5}$ \\
         & 18.0 $-$ 26.0 & ... & ... & 0.33 $\times$ 0.25, $-$53.2 & 120 & ... & ... & ... \\
    G309.92+0.48\small\textsuperscript{b} & 5.5 $-$ 9.0 & 13 50 41.85 & $-$61 35 10.4  & 1.09 $\times$ 0.89, $+$12.3 & 560 & 5.50 & $3.1^{+0.8}_{-0.6} \times 10^{5}$  & $6.6^{+4.8}_{-2.8} \times 10^{5}$ \\
                    & 17.0 $-$ 22.8 & ... & ... & 0.83 $\times$ 0.52, $-$11.1 & 820 & ... & ... & ... \\
    G35.58-0.03\small\textsuperscript{a} & 4.0 $-$ 8.0 & 18 56 22.56 & +02 20 27.7 & 0.32 $\times$ 0.27, $+$64.7 & 8  & 10.2 & $2.4^{+0.3}_{-0.3}\times 10^{5}$ & $4.3^{+2.9}_{-1.7}\times 10^{5}$ \\
             & 18.0 $-$ 26.0 & ... & ... & 0.30 $\times$ 0.29, $+$73.5 & 90 & ... & ... & ... \\
    IRAS 16562-3959\small\textsuperscript{b}\small\textsuperscript{$\dagger$} & 5.0 $-$ 9.0 & 16 59 41 63 & $-$40 03 43.6 & 2.49 $\times$ 1.36, $-$4.13 & 21 & 1.70 & $4.9.0^{+1.9}_{-1.3}\times 10^{4}$ & $2.2^{+2.5}_{-1.2}\times 10^{5}$ \\
              & 17.0 $-$ 19.0 & ... & ... & 0.78 $\times$ 0.39, $+$2.87 & 28 & ... & ... & ...\\
   G305.20+0.21\small\textsuperscript{b} & 5.5 $-$ 9.0 & 13 11 10.49 & $-$62 34 38.8 & 0.97 $\times$ 0.82, $+$86.5 & 70 & 4.10 & $6.2^{+1.7}_{-1.4}\times 10^{4}$ & $2.1^{+2.0}_{-1.0}\times 10^{5}$ \\
          & 17.0 $-$ 22.8 & ... & ... & 0.61 $\times$ 0.40, $-$6.59 & 70 & ... & ... & ... \\
    G49.27-0.34\small\textsuperscript{a} & 4.0 $-$ 8.0 & 19 23 06.61 & +14 20 12.0 & 0.29 $\times$ 0.26, $+$64.3 & 6 & 5.55 & $1.4^{+2.8}_{-0.9}\times 10^{5}$ & $1.0^{+0.5}_{-0.3}\times 10^{5}$ \\
                    & 18.0 $-$ 26.0 & ... & ... & 0.33 $\times$ 0.26, $-$62.2 & 9 & ... & ... & ... \\
    G339.88-1.26\small\textsuperscript{b}\small\textsuperscript{$\ddagger$} & 5.5 $-$ 9.0 & 16 52 04.67 & $-$46 08 34.2 & 2.38 $\times$ 1.44, $-$16.3 & 25 & 2.10 & $3.8^{+1.6}_{-1.1}\times 10^{4}$ & $9.1^{+16.0}_{-5.8}\times 10^{4}$ \\
                    & 17.0 $-$ 22.8 & ... & ... & 0.90 $\times$ 0.34, $-$1.09 & 75 & ... & ... & ... \\
    \enddata
    \tablecomments{ Units of R.A. are hours, minutes, and seconds. Units of Decl. are degrees, arcminutes, and arcseconds. \\
    \textsuperscript{a} Images from the VLA. \\
    \textsuperscript{b} Images from ATCA.\\
    \textsuperscript{c} References cited in \citetalias{Liu_2019}. \\
    \textsuperscript{d} Average and dispersion of the bolometric luminosities of the good models from \citetalias{Telkamp_2025}, these values and other intrinsic properties are reported in Table C1.\\
    \textsuperscript{$\dagger$} Images obtained from Andr\'es Guzm\'an (via public personal archive). Observations reported in \citet{2016ApJ...826..208G}.
    \\
    \textsuperscript{$\ddagger$} Images obtained from Simon Purser (via private communication). Observations reported in \citet{2016MNRAS.460.1039P}.}
\end{deluxetable*}

\begin{deluxetable}{cccc}
    \tabletypesize{\footnotesize}
    \tablewidth{0pt}
    \tablecaption{VLA Calibrators. \label{tab:VLA_Calibrators}}
    \tablehead{
    \colhead{Calibrator} & \colhead{Astrometry Precision\small\textsuperscript{a}} & \colhead{Source Calibrated} & \colhead{Band}}
    \startdata
    J1922+1530 & A & G45.12+0.13     & C \\
    J1924+1540 & A & G45.12+0.13     & K \\
    J1824+1044 & A & G35.58-0.03     & C \\
    J1851+0035 & C & G35.58-0.03     & K \\
    J1922+1530 & A & G49.27-0.34     & C \\
    J1925+2100 & A & G49.27-0.34     & K \\
    \enddata
    \tablecomments{\\
    \textsuperscript{a} Astrometric precisions of A, B and C correspond to positional accuracies of $<$0.002 arcsec, 0.002\textendash0.01 arcsec and 0.01\textendash0.15 arcsec, respectively.}
\end{deluxetable}

\subsection{VLA data}\label{VLA_data}

\subsubsection{The 6 cm data}\label{6cm}

The 6 cm (C-Band) observations were carried out in the A configuration providing angular resolutions $\sim0\farcs3$ and Largest Angular Scale (LAS or $\theta_{LAS}$) of 8.9 arcsec. The project code for the observations of regions G35.58-0.03 and G49.27-0.34 is 18A-294 (PI: V. Rosero) and for region G45.12+0.13 is 19A-216 (PI: V. Rosero). The data consists of two $\sim$2 GHz wide basebands (3 bit samplers) centered at 5.03 and 6.98 GHz. The data were recorded in 30 unique spectral windows (SPWs), each comprised of 64 channels and each channel being 2 MHz wide, resulting in a total bandwidth of 3842 MHz (before “flagging”). Source 3C48 was used as flux density and bandpass calibrator for regions G35.58-0.03 and G49.27-0.34, and 3C286 for region G45.12+0.13.

The data were processed using NRAO’s Common Astronomy Software Applications (CASA) package and we use calibrated data from the VLA Calibration Pipeline version 41154 (Pipeline-CASA51-P2-B) for regions G35.58-0.03 and G49.27-0.34, and the VLA Calibration Pipeline version 42270 (Pipeline-CASA54-P2-B) for region G45.12+0.13. The images were made using the \textit{tclean} task and Briggs \textit{Robust} = 0.5 weighting \citep{1995PhDT.......238B}. Imaging of regions G35.58-0.03  and G45.12+0.13 also included one round of phase-only self calibration using solution intervals ranging from 30s to 60s depending on frequency.

For all the regions we made two images, each of a $\sim$2 GHz baseband composed of 15 SPWs, and also a combined image using data from both basebands with a total of 30 SPWs. All maps were primary beam-corrected. Columns 5 and 6 of Table \ref{tab:SOMA_Sources} list the synthesized beam (size and position angle) and the rms of the combined images.

Table \ref{tab:VLA_Phase_Times} shows approximations of the observation times for each region at both 6 and 1.3 cm bands. All the observations were made alternating between the target source and the phase calibrator. Columns 3 and 4 show these alternating times, and column 5 shows the total on source integration time for each observation.

\subsubsection{The 1.3 cm data}\label{1.3 cm}

The 1.3 cm (K-Band) observations were made in the B configuration providing angular resolutions $\sim0\farcs4$ and Largest Angular Scale (LAS or $\theta_{LAS}$) of 7.9 arcsec. The project code for all  observations in this band is 19A-216 (PI: V. Rosero). The data consists of two $\sim$4 GHz wide basebands (3 bit samplers) centered at 20.4 and 24.4 GHz. The data was recorded in 60 unique SPWs, comprised of 64 channels and each channel being 2 MHz wide, resulting in a total bandwidth of 7680 MHz, before “flagging”. Source 3C286 was used as flux density and bandpass calibrator. The data reduction was done in the same manner as that for the C-Band observations, using the VLA Calibration Pipeline version 42270 (Pipeline-CASA54-P2-B) for all the regions.

The images were made using the \textit{tclean} task and Briggs \textit{Robust} = 0.5 weighting. We made two images, each of a $\sim$4 GHz baseband composed of 30 SPWs, and also a combined image using data from both basebands with a total of 60 SPWs. All maps were primary beam-corrected. Columns 5 and 6 of Table \ref{tab:SOMA_Sources} show the synthesized beam (size and position angle) and the rms of the combined images.

\begin{deluxetable}{ccccc}
    \tabletypesize{\footnotesize}
    \tablewidth{0pt}
    \tablecaption{Observation times for sources observed using the VLA. \label{tab:VLA_Phase_Times}}
    \tablehead{\colhead{Source} & \colhead{Band} & \colhead{Target} & \colhead{Phase} & \colhead{Total}\\
    \colhead{} & \colhead{} & \colhead{Source\textsuperscript{a}} & \colhead{Calibrator\textsuperscript{a}} & \colhead{Integration}\\
    \colhead{} & \colhead{} & \colhead{(min)} & \colhead{(min)} & \colhead{(min)}}
    \startdata
    G45.12+0.13 & C & 8.4 & 0.45 & 42 \\
         & K & 1.2 & 0.54 & 23 \\
    G35.58-0.03 & C & 8.2 & 0.52 & 41 \\
         & K & 2.5 & 0.62 & 23 \\
    G49.27-0.34 & C & 8.6 & 0.65 & 43 \\
         & K & 2.3 & 0.51 & 20 \\
    \enddata
\tablecomments{  \\
    \textsuperscript{a} Alternating time from source to calibrator. }
\end{deluxetable}

\subsection{ATCA data}\label{ATCA_Data}
For regions G309.92+0.48 and G305.20+0.21, we used our own ATCA observations; meanwhile, for regions IRAS 16562-3959 and G339.88-1.26, we used ATCA observations reported by \cite{2016ApJ...826..208G} and \cite{2016MNRAS.460.1039P}. 

In the following, we describe our new ATCA observations. The project code of the observations for the regions G309.92+0.48 and G305.20+0.21 is C3396 (PI: V. Rosero). These observations were made using ATCA in the 6A and 6B configurations in October 2020. Observations were made at four frequencies: 5.5, 9.0, 17.0 and 22.8 GHz, each with a bandwidth of 2048 MHz (XX, YY, XY, and YX polarizations) that was split into channels of width 1 MHz. 

Individual scan times on flux and phase calibrators were 5 minutes and 90 seconds respectively and on target sources were between 3 and 45 minutes dependent on atmospheric conditions and observing frequencies. The average total integration time on each science target was 75 minutes. The bandpass and flux density was calibrated with source B1934−638 using the Stevens-Reynolds 2016 model \citep{2016ApJ...821...61P}. The ATCA data were calibrated in CASA following standard interferometric procedures for flagging, flux-density transfer and phase referencing. Imaging of region G305.20+0.21 also included one round of phase-only self calibration using solution intervals ranging from 30s to 60s depending on frequency.

For the two regions observed with project code C3396 we made four images centered around the four frequencies previously described. However, similarly to the VLA observations, we also made two additional images with the combined data at the lowest frequency observations (5.5 and 9.0 GHz) and at the highest frequency observations (17 and 22.8 GHz). This resulted in two frequency-combined radio maps centered around 7.2 and 19.9 GHz respectively.

For region IRAS 16562-3959 we used the data from \cite{2016ApJ...826..208G} at 5, 9, 17 and 19 GHZ: more information about the observations can be found in section 2 of their paper. Meanwhile, for region G339.88-1.26 we used data from  \cite{2016MNRAS.460.1039P} at 5.5, 9.0, 17.0 and 22.8 GHz: more information about the observations can be found in section 3.1 of their paper. In order to have frequencies comparable to the combined images made for the other sources, we select the lowest and highest frequency maps of both regions for use in the direct comparison with the combined maps, but we used the images in the same way that they were published.

\section{Results}\label{subsec:results}

Following the methods of \citetalias{Rosero_2019}, we consider a radio source being detected when the peak intensity $I_\nu$ is at least 5 times the rms ($\sigma$) in either of the baseband-combined images (see Section \ref{VLA_data} and \ref{ATCA_Data}) at the different frequencies. For non-detections in the combined images, we report an upper limit value of 3$\sigma$ for the flux density at the given frequency.

Figure \ref{fig:VLA_Contours} shows the radio contour plots of the lower frequency in red (C-band (6 cm) for the VLA observations and 5, 5.5 or 7.2 GHz for ATCA observations) and the higher frequency in cyan (K-band (1.3 cm) for the VLA observations and 17, 19 or 19.9 GHz for ATCA observations) for all the radio sources detected in our sample and overlaid to SOFIA-FORCAST 37 $\mu$m images (from \citetalias{Liu_2019}). 

The blue circles represent the SOMA aperture set by \citetalias{Fedriani_2023} using the automated algorithm to choose the optimal aperture size (see Table C1 of \citetalias{Telkamp_2025}), except for IRAS 16562-3959 N, where the aperture radius is set from the 37.1 $\mu$m image reported by \citetalias{Liu_2020} (see Table 2 of \citetalias{Liu_2020}), this aperture size was used to build the IR SEDs. The infrared images presented in \citetalias{Liu_2019} (including SOFIA, Herschel, and Spitzer data) and the VLA data presented here have astrometric accuracies better than 1.5$^{\prime\prime}$ and 0$^{\prime\prime}$.1 (see Table \ref{tab:VLA_Calibrators}), respectively. Meanwhile, the ATCA data presented here have astrometric accuracies better than 0$^{\prime\prime}$.4, including both the phase calibrator's positional uncertainty and errors in transferring phases to the target.

\begin{figure*}[ht!]
\figurenum{1}
\begin{center}
\includegraphics[width=0.49\linewidth]{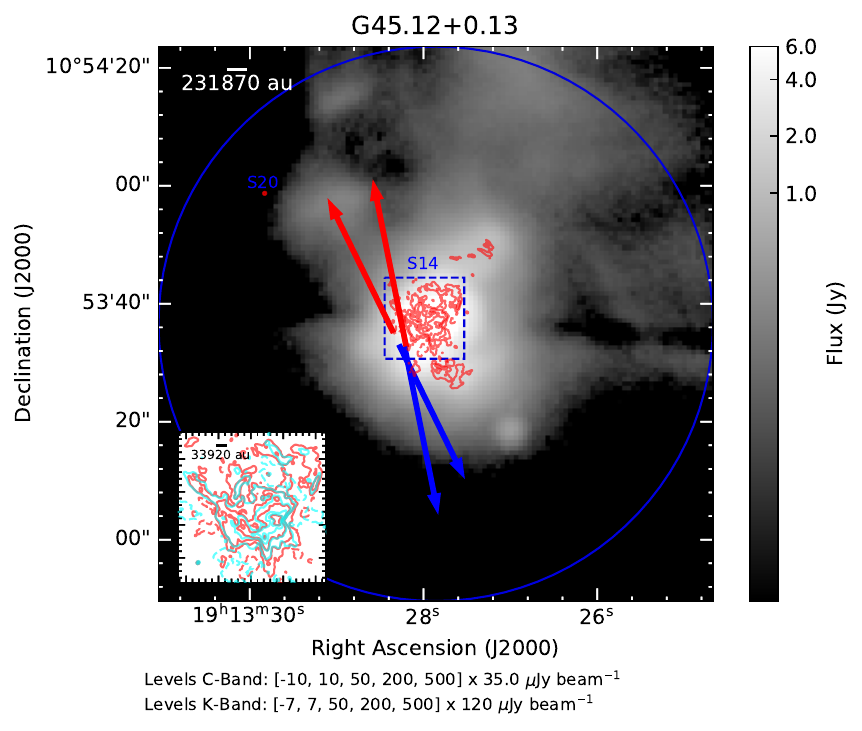}\quad\includegraphics[width=0.49\linewidth]{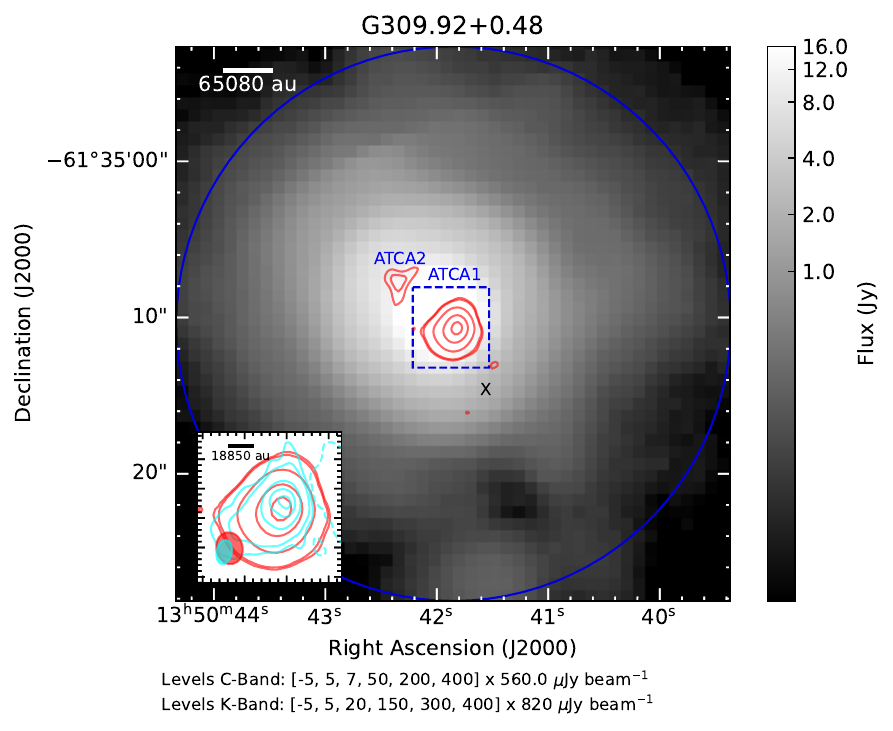} \\
\includegraphics[width=0.49\linewidth]{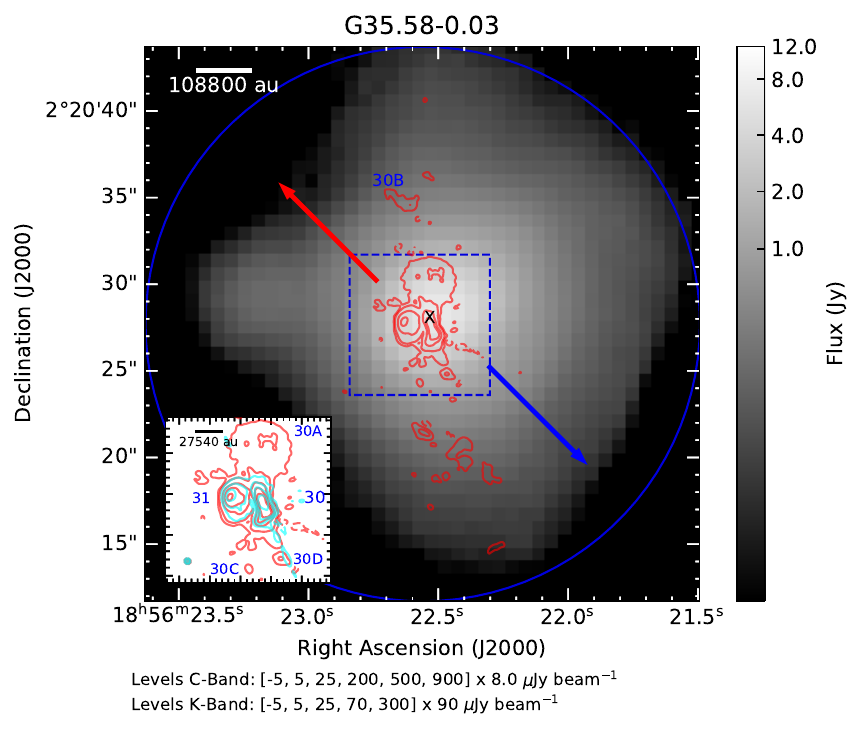}\quad\includegraphics[width=0.49\linewidth]{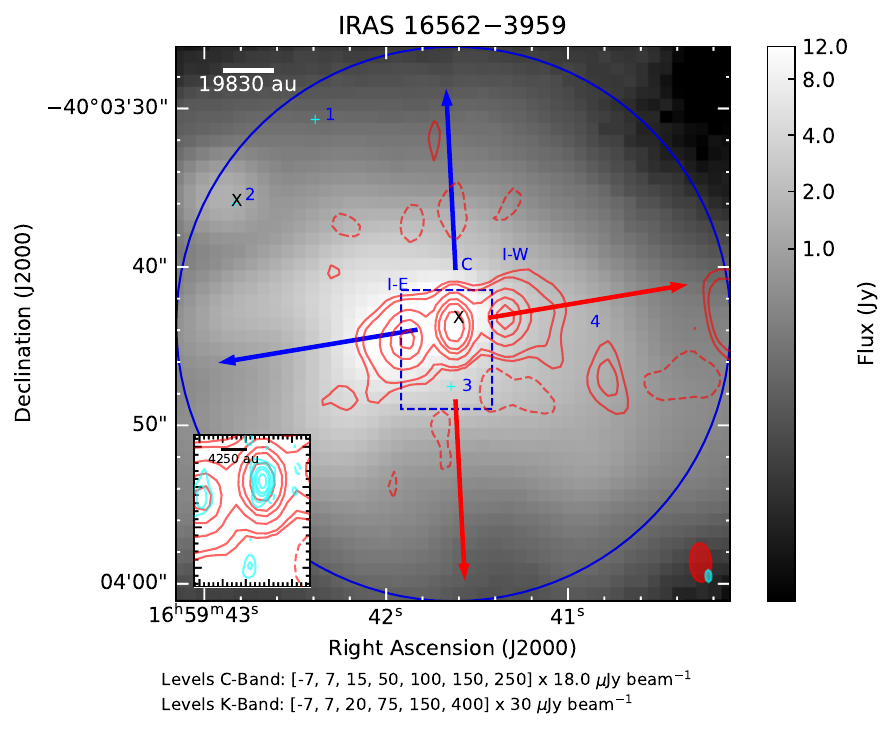} \\
\caption{Images are SOFIA-FORCAST 37 $\mu$m (from \citetalias{Liu_2019}) with VLA contours $-$ red: lower frequency (C-band (6 cm) for the VLA observations and 5, 5.5 or 7.2 GHz for ATCA observations); cyan: (K-band (1.3 cm) for the VLA observations and 17, 19 or 19.9 GHz for ATCA observations) $-$ of the combined radio maps overlaid. The centimeter emission observations for regions IRAS 16562--3959 and G339.88--1.26 are originally presented in \citet{2016ApJ...826..208G} and \citet{2016MNRAS.460.1039P}, respectively.
The cyan crosses in IRAS 16562--3959 denotes the position of detections only at the high frequency bands (cyan contours).
The black $\times$ denotes the position of the mm core on the regions with millimeter observations (references as follows: G309.92+0.48: \cite{2005MNRAS.363..405H}, G35.58-0.03: \cite{2014ApJ...784..107Z},
IRAS 16562 3959: \cite{2014ApJ...796..117G}, G305A: \cite{2006MNRAS.365..321W} and G339.88-1.26: \cite{2019ApJ...873...73Z}). 
The blue dashed squares correspond to the area of the inset image showing a zoom-in of the central region, and the synthesized beams are shown in the lower corners of these insets.
The blue circles are the SOMA apertures used by \citetalias{Telkamp_2025} and reported in Table C1.
The aperture radius is defined using the optimal aperture size algorithm developed by \citetalias{Fedriani_2023}. 
The blue and red arrows represent the direction of a molecular outflow detected toward the region, only in regions with known molecular outflows (more details in Section \ref{Morph}). A scale bar in units of au is shown in the upper left of the figures. \label{fig:VLA_Contours}}
\end{center}
\end{figure*}

\begin{figure*}[ht!]
\figurenum{1}
\begin{center}
\includegraphics[width=0.49\linewidth]{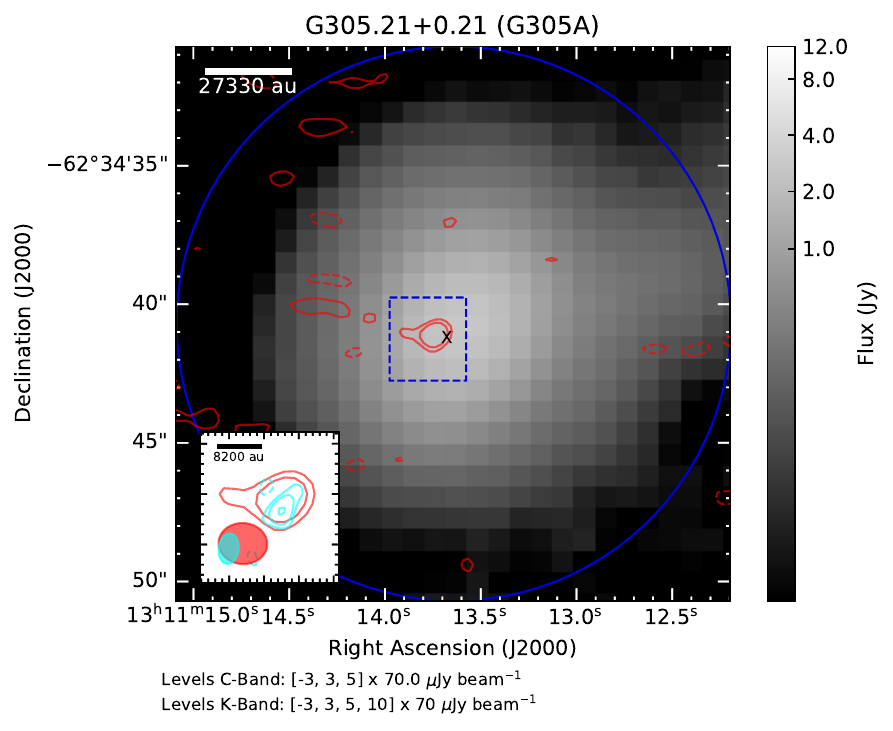}\quad\includegraphics[width=0.49\linewidth]{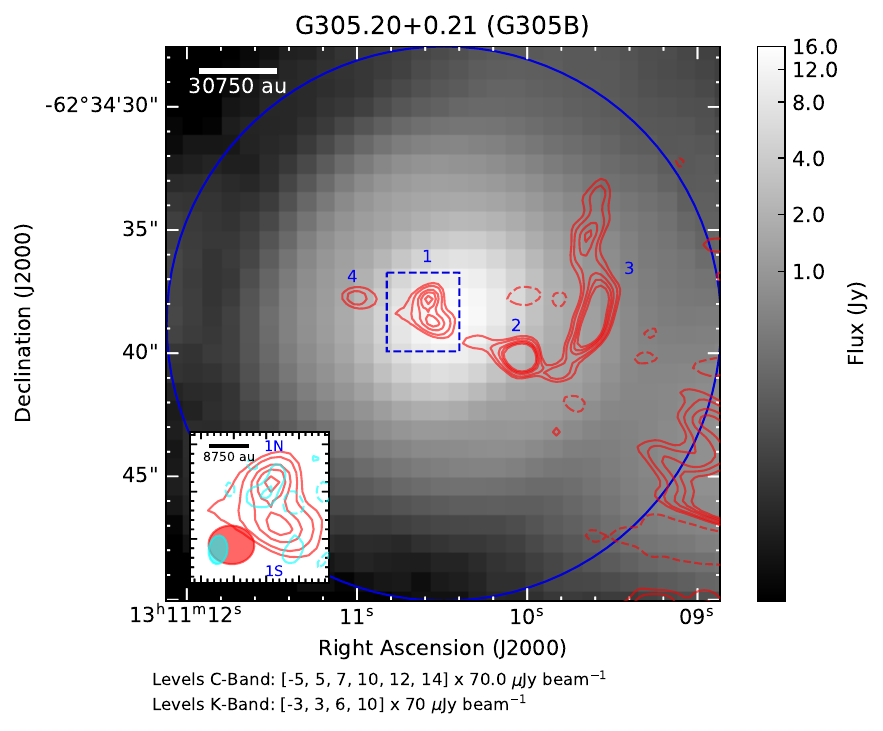} \\
\includegraphics[width=0.49\linewidth]{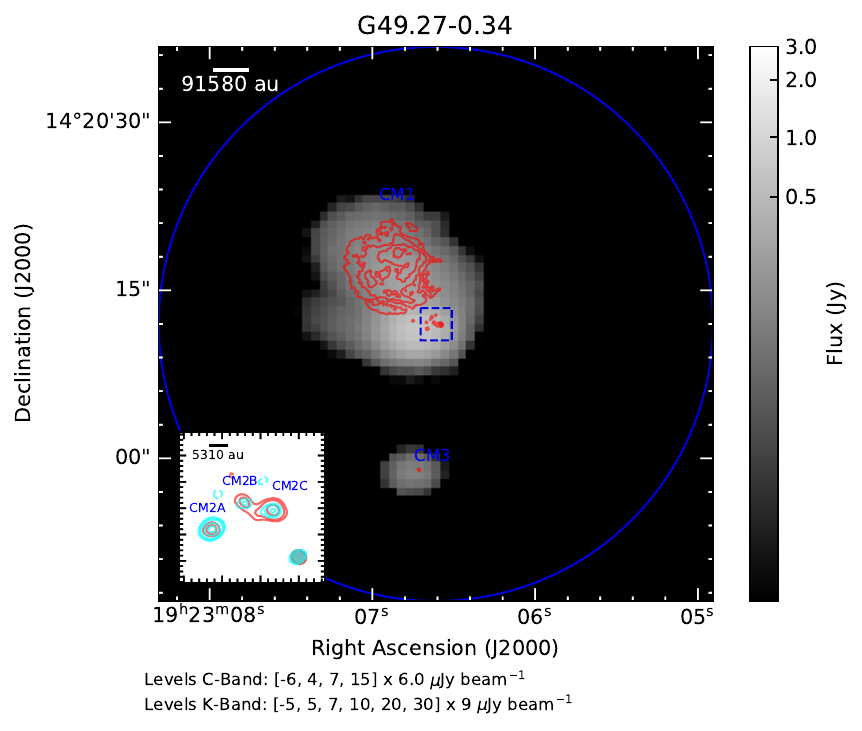}\quad\includegraphics[width=0.49\linewidth]{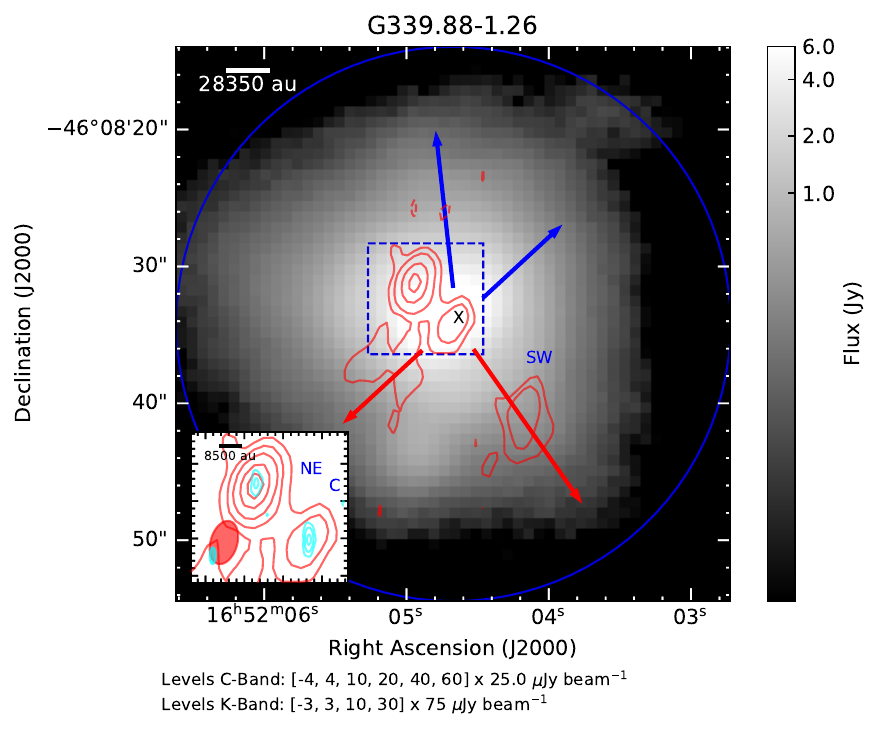} \\
\caption{(Continued)}
\end{center}
\end{figure*}

Table \ref{tab:Parameters_Radio} reports the radio parameters for each of the nine protostars studied in this paper. In line with the methods developed in \citetalias{Rosero_2019}, these parameters were measured based on different size scales, as follows.
The SOMA scale refers to the size of the aperture radius used in \citetalias{Liu_2019} and \citetalias{Telkamp_2025} to measure IR fluxes. The \textit{Intermediate} scale is based on the morphology of the radio source, specifically whether the detections appear to be of jet-like nature. 
The \textit{Inner} scale is given by the size of the central radio detection that is likely most closely associated with the driving protostar, as set by the SOFIA data and the center of the infrared SEDs analyzed in \citetalias{Liu_2019} (i.e., association with compact millimeter dust continuum emission). 
For the VLA data presented, we determined the flux density $S_\nu$ in each wideband image in the SOMA and \textit{Intermediate} scales by using the task \textit{imstat} of CASA, either enclosing the SOMA aperture in a circular region or enclosing the elongated jet-like structure in a box, respectively (see Table \ref{tab:SOMA_scales}). The uncertainties of the flux densities for these two scales are estimated as $\sigma_{\mathrm{image}}$(npts/beam area)$^{0.5}$ added in quadrature with an assumed 10\% error in calibration, where $\sigma_{\mathrm{image}}$ is the rms of the image, npts is the number of pixels enclosed in the box or the circular region, and the beam area is the number of pixels within a synthesized beam of the image. For the \textit{Inner} scale, we determined the flux density using the task \textit{imfit} of CASA, and the uncertainty was estimated using the statistical error from the Gaussian fit added in quadrature with an assumed 10\% error in calibration.

Column 1 and 2 of Table \ref{tab:Parameters_Radio} shows the region and the given scale; for each, columns 3 and 4 report the R.A. and Decl. For the SOMA scale, these refer to the pointing center observations of SOFIA-FORCAST (see \citetalias{Liu_2019} Table 1); for the \textit{Intermediate} scale, they refer to a middle point in the jet-like detection; and for the \textit{Inner} scale, they refer to the R.A. and Decl. of the peak intensity of the central detected object. The following columns are the flux densities ($S_\nu$) at different frequencies, with the uncertainties given in parentheses (for non-detections, the upper limits denotes the minimum detectable flux of a point source). The last column in Table \ref{tab:Parameters_Radio} reports the spectral indices and their uncertainties at each scale (see Section \ref{Radio_SEDs}). Since we are limited by the Large Angular Scale (LAS), the radio data are not sensitive to extended emission over scales as large as the SOMA and possibly the \textit{Intermediate} scales, the flux measurements represent the sum over all compact sources within the scale, and the spectral indices contain some corresponding uncertainty (more details in Section \ref{Radio_SEDs}). The error bars for these measurements are large, due to having many independent beams within the scale. Consequently, the uncertainties in the flux density and spectral index at these scales tend to be higher for some sources. It is important to note, however, that smaller structures such as the \textit{Inner} scale are less affected.

\startlongtable
\begin{deluxetable*}{ccccccccc}
    \tabletypesize{\scriptsize}
    \tablewidth{0pt}
    \tablecaption{Parameters from Radio Continuum. \label{tab:Parameters_Radio}}
    \tablehead{
    \colhead{Region} & \colhead{Scale} & \colhead{R.A} & \colhead{Decl.} & \colhead{$S_{5.0}$ $_{\text{GHz}}$} & \colhead{$S_{7.0}$ $_{\text{GHz}}$} & \colhead{$S_{20.4}$ $_{\text{GHz}}$} & \colhead{$S_{24.4}$ $_{\text{GHz}}$} & \colhead{Spectral} \\
    \colhead{} & \colhead{} & \colhead{(J2000)} & \colhead{(J2000)} & \colhead{(mJy)} & \colhead{(mJy)} & \colhead{(mJy)} & \colhead{(mJy)} & \colhead{Index}}
    \startdata
    & SOMA & 19:13:27.86 & +10:53:36.60 & 1797.44(180.6) & 1855.93(186.5) & 3673.00(367.5) & 3593.19(360.71) & 0.5(0.1) \\
    G45.12+0.13 & \textit{Intermediate} & \nodata & \nodata & \nodata & \nodata & \nodata & \nodata & \nodata \\
    & \textit{Inner} & 19:13:27.87 & +10:53:36.17 & 1702.69(51.11) & 1928.91(57.89) & 3668.96(110.0) & 3722.08(111.7) & 0.5(0.1) \\
    \hline 
    & SOMA & 13:50:41.85 & $-$61:35:10.40 & 454.76(47.4) & 753.47(79.0) & 1017.68(108.6) & 955.54(115.6) & 0.5(0.1) \\
    G309.92+0.48\textsuperscript{a} & \textit{Intermediate} & \nodata & \nodata & \nodata & \nodata & \nodata & \nodata & \nodata \\
    & \textit{Inner} & 13:50:41.84 & $-$61:35:10.85 & 428.18(14.6) & 713.91(29.3) & 1085.90(57.86) & 1191.15(76.69) & 0.7(0.1) \\
    \hline
    & SOMA & 18:56:22.52 & +02:20:27.20 & 172.38(17.3) & 199.20(20.0) & 302.12(30.6) & 321.61(33.5) & 0.4(0.1) \\
    G35.58-0.03 & \textit{Intermediate} & \nodata & \nodata & \nodata & \nodata & \nodata & \nodata & \nodata \\
    & \textit{Inner \textsuperscript{$\dagger$}} & 18:56:22.52 & +02:20:30.53 & 84.00(27.32) & 118.00(35.05) & 237.00(91.14) & 263.00(101.47) & 0.7(0.2) \\
    \hline
    & SOMA & 16:59:41.63 & $-$40:03:43.60 & 18.92(1.9) & 23.50(2.4) & 34.27(3.7) & 31.56(3.5) & 0.4(0.1) \\
    IRAS 16562-3959\textsuperscript{b} & \textit{Intermediate} & 16:59:42.17 & $-$40:03:46.85 & 19.66(2.0) & 27.63(2.8) & 31.84(3.2) & 33.34(3.4) & 0.4(0.1) \\
    & \textit{Inner} & 16:59:41.63 & $-$40:03:43.73 & 8.18(0.87) & 13.40(1.43) & 20.80(2.19) & 22.90(2.43) & 0.8(0.1) \\
    \hline
    & SOMA & 16:59:43.01 & $-$40:03:11.56 & $<$0.39 & 0.47(0.22) & 0.59(0.64) & 0.87(0.66) & $<$0.7 \\
    IRAS 16562-3959 N\textsuperscript{b} & \textit{Intermediate} & \nodata & \nodata & \nodata & \nodata & \nodata & \nodata & \nodata \\
    & \textit{Inner} & \nodata & \nodata & $<$0.06 & $<$0.06 & $<$0.09 & $<$0.08 &  $<$0.4 \\
    \hline
    & SOMA & 13:11:13.64 & $-$62:34:40.70 & 1.25(2.37) & 1.05(2.14) & 0.71(2.18) & 3.22(8.43) & -0.3(3.1) \\
    G305.21+0.21\textsuperscript{a} & \textit{Intermediate} & \nodata & \nodata & \nodata & \nodata & \nodata & \nodata & \nodata \\
    (G305A) & \textit{Inner} & 13:11:13.75 & $-$62:34:41.31 & 0.96(0.01) & 0.74(0.04) & 0.91(0.09) & 2.70(0.27) & 0.2(0.2) \\
    \hline
    & SOMA & 13:11:10.49 & $-$62:34:38.80 & 16.57(2.11) & 12.98(2.36) & 3.42(2.63) & $<$29.72  & $<$-1.0 \\
    G305.20+0.21\textsuperscript{a} & \textit{Intermediate} & 13:11:10.52 & $-$62:34:39.08 & 7.23(0.83) & 4.55(0.77) & 1.59(0.83) & $<$9.33  & $<$-1.2 \\
    (G305B) & \textit{Inner} & 13:11:10.56 & $-$62:34:38.31 & 3.39(0.13) & 1.61(0.89) & 1.37(0.19) & 1.92(0.38) & -0.7(0.2) \\
    \hline
    & SOMA & 19:23:06.61 & +14:20:12.00 & 44.30(4.54) & 52.93(5.44) & 26.11(3.34) & 40.86(4.50) & -0.3(0.1) \\
    G49.27-0.34\textsuperscript{c} & \textit{Intermediate} & 19:23:06.69 & +14:20:11.22 & 0.21(0.04) & 0.01(0.04) & 0.33(0.08) & 0.34(0.07) & 0.7(0.5) \\
    & \textit{Inner} & 19:23:06.66 & +14:20:11.66 & 0.04(0.01) & 0.03(0.01) & 0.27(0.02) & 0.35(0.02) & 1.6(0.2) \\
    \hline
    & SOMA & 16:52:04.67 & $-$46:08:34.20 & 5.50(0.71) & 3.64(0.68) & 4.11(2.16) & 4.27(4.61) & -0.6(0.5) \\
    G339.88-1.26\textsuperscript{a} & \textit{Intermediate} & 16:52:04.62 & $-$46:08:36.04 & 5.31(0.57) & 6.86(0.73) & 4.58(1.04) & 4.03(2.05) & -0.1(0.2) \\
    & \textit{Inner} & 16:52:04.94 & $-$46:08:31.32 & 0.86(0.09) & 1.51(0.16) & 2.44(0.26) & 3.00(0.33) & 0.9(0.1)\\
    \enddata
    \tablecomments{The Intermediate scale corresponds to the extent of the radio jet (candidate). Units of R.A. are hours, minutes, and seconds. Units of Decl. are degrees, arcminutes, and arcseconds. \\
    \textsuperscript{$\dagger$} Inner scale corresponds to source 30 that we are resolving into sources 30N and 30S (see section \ref{g35section} and Table \ref{tab:Multiplicity_G35} for more details). \\
    The following are ATCA observations: \\
    \textsuperscript{a} The frequencies at which the flux was measured were 5.5, 9.0, 17.0 and 22.8 GHz. \\
    \textsuperscript{b} The frequencies at which the flux was measured were 5.0, 9.0, 17.0 and 19.0 GHz. \\
    \textsuperscript{c} The frequencies at which the flux was measured were 5.0, 7.0, 20.4 and 24.4 GHz.}
\end{deluxetable*}

\begin{deluxetable}{cccc}
    \tabletypesize{\footnotesize}
    \tablewidth{0pt}
    \tablecaption{Size scales of flux measurements. \label{tab:SOMA_scales}}
    \label{tab:Sources_Scales}
    \tablehead{
    \colhead{Region} & \colhead{SOMA} & \colhead{Intermediate} & \colhead{Inner}\\
    \colhead{} & \colhead{$R$ ($^{\prime\prime}$)} & \colhead{\textit{w($^{\prime\prime}$) $\times$ h($^{\prime\prime}$)}} & \colhead{\textit{a($^{\prime\prime}$) $\times$ b($^{\prime\prime}$)}}}
    \startdata
    G45.12+0.13       & 47.0   & \nodata & 3.67 $\times$ 2.33\small\textsuperscript{a}\\
    G309.92+0.48      & 17.75  & \nodata & 1.96 $\times$ 1.75\small\textsuperscript{a}\\
    G35.58-0.03       & 16.0  & \nodata & 2.88 $\times$ 0.50\small\textsuperscript{a}\\
    IRAS 16562-3959   & 17.5   & 13.2 $\times$ 6.82 & 2.46 $\times$ 1.61\small\textsuperscript{a}\\
    IRAS 16562-3959 N & 7.7   & \nodata & \nodata\\
    G305.20+0.21 A    & 10.0   & \nodata & 1.16 $\times$ 1.11\small\textsuperscript{a}\\
    G305.20+0.21      & 11.25   & 8.68 $\times$ 4.45 & 1.12 $\times$ 0.77\small\textsuperscript{a}\\
    G49.27-0.34       & 24.75   & 2.01 $\times$ 1.14 & 0.44 $\times$ 0.37\small\textsuperscript{b}\\
    G339.88-1.26      & 20.25   & 13.8 $\times$ 17.9 & 2.38 $\times$ 0.91\small\textsuperscript{a}\\ 
    \enddata
    \tablecomments{The reported values correspond to a circle of radius \textit{R} for the SOMA scale and a box of height \textit{h} and width \textit{w} for the \textit{Intermediate} scale. In some cases, the inner scales of the C- and K-band images differ. In these situations, we report the larger area and consider any emission within the inner region. We also distinguish between two types of regions: \\
    \textsuperscript{a} Scale corresponds to image component (convolved with beam) size from the task imfit, of major axis \textit{a} and minor axis \textit{b}. \\
    \textsuperscript{b} Scale corresponds to an ellipse of major axis \textit{a} and minor axis \textit{b}.}
\end{deluxetable}

\subsection{Morphology and Multiplicity}\label{Morph}

All the target regions presented in this paper have been observed in the centimeter continuum. 
Our criteria for multiplicity is the presence of two or more spatially resolved 5$\sigma$ detections. Furthermore, we describe source morphology as ``compact'' if the detection shows no structure on the scale of a few number of synthesized beams or greater. Otherwise we describe it as ``extended''. Below, we describe the centimeter wavelength detections toward each target; for a detailed literature review on each of these regions, see \citetalias{Liu_2019}.

\subsubsection{G45.12+0.13}

G45.12+0.13 is a massive star-forming region, also know as IRAS 19111+1048, that is associated with  an UC HII region. It is located at 7.4~kpc \citep{2011ApJ...736..149G} and is part of the Galactic Ring Survey Molecular Cloud (GRSMC) G45.46+0.05, a large star formation complex \citep{1989ApJS...69..831W, 2001ApJ...551..747S}.

G45.12+0.13 has been analyzed at multiple wavelengths ranging from IR, sub-mm and radio \citep{1997ApJ...478..283H,2006ApJ...637..400V,2022PASA...39...24A, 2021A&A...645A.110Y}. \cite{1997ApJ...478..283H} reported bipolar outflows towards this region using CO ($J = 6–5$) maps. From this study, the highest velocity outflow is centered around a source termed S14, while an additional outflow is centered around another UC HII region source G45.12+0.13 West. Both of these UC HII regions contain type-I OH masers \citep{2000ApJS..129..159A}. Additionally, with radio observations from the Giant Metrewave Radio Telescope (GMRT) at 1280, 610 and 325 MHz, \cite{2006ApJ...637..400V} suggested that this region shows a highly homogeneous ionized medium morphology and \citetalias{Liu_2019} confirm this morphology with their MIR images. Furthermore, \cite{2006ApJ...637..400V} concluded that G45.12+0.13 contains a cluster of ZAMS stars energizing a compact and evolved HII region. In total they reported 20 sources, with S14 being the central UC HII source and S20 a non-thermal source according to its radio emission.

\begin{figure*}[ht!]
\figurenum{2}
\begin{center}
\includegraphics[width=0.85\linewidth]{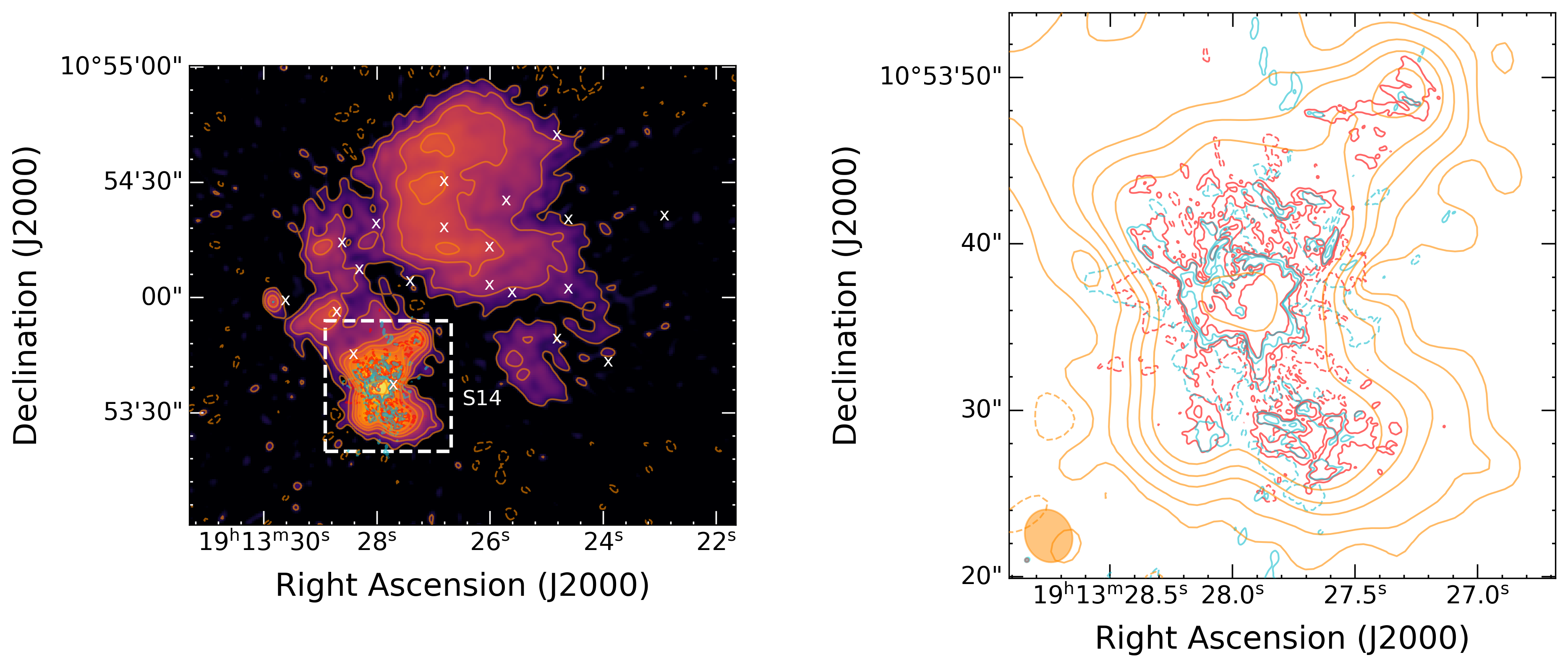}
\caption{Left is L-band VLA (20 cm) continuum maps with VLA contours overlaid (1.3 cm--cyan, 6 cm--red, 20 cm--blue) of G45.12+0.13. The white $\times$ symbol are the location of the sources in the cluster reported by \cite{2006ApJ...637..400V}. Note the offset
in the peak intensity position from the results of \cite{2006ApJ...637..400V} with respect to the VLA observations. Right is showing a zoom region of source S14 with the VLA contours of the three bands. The contour levels are at [-5, 5, 20, 100] $\times$ 35 $\mu$Jy/beam for C-band, at [-5, 5, 15] $\times$ 120 $\mu$Jy/beam for K-band and at [-3, 3, 10, 30, 55, 100, 300, 500] $\times$ 0.25 mJy/beam for the L-band. The synthesized beams  are shown in the lower left corner. \label{fig:G45_Lband}}
\end{center}
\end{figure*}

From our VLA observations at 1.3 and 6~cm we report two detections corresponding to sources S14 and S20 from \cite{2006ApJ...637..400V}. Both detections have an offset in the peak intensity position from the results of \cite{2006ApJ...637..400V} using GMRT observations, by a difference of around 2.1$^{\prime\prime}$ to the SE (see Figure \ref{fig:G45_Lband}). This systematic offset may be due to astrometric uncertainties in the GMRT data which we do not have enough information to quantify. Contributions to the uncertainty are expected to include calibrator position, phase transfer and self-calibration. The peak can also change with resolution given the presence of extended components.

It is important to highlight that despite our 
data being of higher resolution, we only detect two of the twenty compact sources from the cluster reported by \cite{2006ApJ...637..400V}. From our data we infer that most of the emission to the northern side of the central UC HII region arises from an extended and diffuse irregular cloud of ionized material instead of a cluster of compact sources. The presence of a position offset does not change this conclusion.

In order to confirm our results, we produce an image using shorter baseline observations from the VLA archive (project code 20A-519; PI: David Neufeld) at L-band (1.5 GHz) in order to compare the results with the 1.28 GHz image from \cite{2006ApJ...637..400V}. Figure \ref{fig:G45_Lband} shows this L-band continuum image with overlaid C, K and L-band VLA contours. The positions of the cluster of point sources reported by \cite{2006ApJ...637..400V} at 1280~MHz are marked with a white $\times$ symbols.

Based on these results, we clearly see that from our images at 1.3 and 6 cm (see Figure \ref{fig:VLA_Contours}), the diffuse ionized material has been filtered out by the long baselines, hence we only detect the very compact point source S20 and the most inner part of the UC HII region source S14. Furthermore, from the shorter baseline image we infer that most of the emission to the north west of the central UC HII region arises from an extended and diffuse irregular cloud of ionized material instead of a cluster of compact sources.

From our VLA observations, source S14 is located at R.A.(J2000) $=19^{h}13^{m}27^{s}.87$, Decl.(J2000)$=+10^{\circ}53^{\prime}36.^{\prime\prime}2$ with a spectral index of 0.5. S14 presents a very complex morphology, and in fact at the shorter frequency (L-band) it appears to have an hour-glass morphology, as seen in the right plot of Figure \ref{fig:G45_Lband}. This observed morphology is similar to the one detected towards source G45.47+0.05 by \citet{2019ApJ...886L...4Z}, who interpreted this as arising from a photoionized outflow. Furthermore, the molecular outflows observed towards G45.12+0.13 (see Figure \ref{fig:VLA_Contours}) are consistent with the direction of the axis of the hour-glass of source S14. 

Meanwhile S20 is located at R.A.(J2000) $=19^{h}13^{m}29^{s}.83$, Decl.(J2000) $=+10^{\circ}53^{\prime}58.^{\prime\prime}7$, presents a very compact morphology at both frequency bands and a spectral index of $-$0.9, a value that is consistent with the non-thermal detection made by \cite{2006ApJ...637..400V}. Moreover, \cite{2006ApJ...637..400V} concluded that, even though source S20 is in the vicinity of G45.12+0.13, it is not likely to be associated with the UCHII region. If this source is indeed independent, the nature of this source could be extragalactic, but more information and observations to confirm this assessment are needed.

\subsubsection{G309.92+0.48}

G309.92+0.48 or IRAS 13471-6120 is a region located at 5.5 kpc \citep{2010MNRAS.405.1560M}. It is associated with OH and methanol masers, as well as radio continuum emission from an UC HII region at several radio observations ranging from 5 to 24 GHz
\citep{1997MNRAS.289..203C, 1998MNRAS.300.1131P, 1998MNRAS.301..640W, 2025MNRAS.538.2267P}. \cite{2021A&A...645A.110Y, 2022A&A...658A.160Y} reported a CO molecular outflow towards this region. In the MIR, G309.92+0.48 is resolved into several sources: \cite{2000ApJS..130..437D} reported it to be composed of a total of 6 point sources (labeled 1 to 6) using high resolution ($\sim0.3$$^{\prime\prime}$) 11.7~$\mu$m data from \textit{Gemini} (see also \citetalias{Liu_2019}).

From our ATCA observations, we have detected two sources that we name ATCA1 and ATCA2 (see Figure \ref{fig:VLA_Contours}). Source ATCA1 is detected at all frequencies (5.5, 9, 17, and 22.8 GHz) and source ATCA2 is only detected at the lower frequencies (5.5 and 9 GHz). Both sources are consistent with the previously reported sources by \citet{1998MNRAS.300.1131P} at 8.6 GHz, sources that are associated with the MIR sources 1 and 2. \citet{2010MNRAS.405.1560M} presented radio continuum observations at 20 GHz towards this region, detecting only source ATCA1, which is consistent with our results of non-detection for source ATCA2 at higher frequencies.

ATCA1 is a slightly elongated source located at R.A.(J2000) $=13^\textnormal{h}50^\textnormal{m}41^\textnormal{s}.84$, Decl.(J2000) $=-61^{\circ}35^{\prime}10.^{\prime\prime}8$ and we report a spectral index of 0.7. ATCA2 is located at R.A.(J2000) $=13^\textnormal{h}50^\textnormal{m}42^\textnormal{s}.29$, Decl.(J2000) $=-61^{\circ}35^{\prime}07.^{\prime\prime}7$ with a spectral index of $-$0.7. Both of these sources show a similar morphology to the previous radio observations mentioned above.

For ATCA1, \citet{2010MNRAS.405.1560M} reported a spectral index of 1.2 between 0.843 and 20 GHz, and \citet{2025MNRAS.538.2267P} reported a spectral index of 0.5 between 5 and 24 GHz. Our estimated value ($\sim$0.7) is similar to the one reported by \citet{2025MNRAS.538.2267P} at similar frequencies. The estimated spectral index value is consistent with either a typical UC H II region \citep{2018A&A...615A.103K} or collimated ionized jets \citep{Reynolds1986Continuum, 2016MNRAS.460.1039P, 2021MNRAS.504..338P}. Based on our results, we classify this source as an optically thick thermal emitter. However, we cannot rule out the possibility of a radio jet, since masers and molecular outflows have been reported toward this source, and a nearby non-thermal component is present. ATCA2, on the other hand, exhibits non-thermal emission, with an upper limit of $\alpha < -0.7$, and shows slight extension, as seen in Fig.~\ref{fig:VLA_Contours}. The nature of this source remains uncertain: given the large distance to this region and the observed morphology, an extragalactic origin cannot be ruled out. Alternatively, this source could represent a jet lobe if ATCA2 was indeed a radio jet. Additional information is required to make a more definitive assessment of the nature of this source.

\subsubsection{G35.58-0.03}\label{g35section}

G35.58−0.03 (hereafter G35.58) is a star-forming region located at a kinematic distance of 10.2 kpc \citep{WatsonG35}. G35.58 has been resolved into two UC HII regions (from now on called sources 30 and 31) using 2 and 3.6~cm observations, where water and OH masers have been detected towards source 30 \citep{1994ApJS...91..659K, 1995MNRAS.272...96C}. Additionally, \cite{2014ApJ...784..107Z} shows evidence of an ionized outflow driving a NE-SW molecular outflow centered in source 30. 

With high-resolution 11.7 $\mu$m \textit{Gemini} observations, \citetalias{Liu_2019} resolved the MIR emission into a main bright peak with two extended diffuse emissions to the north and the northwest. The main bright peak is associated with the eastern UC HII region (source 31). 

We present 6 and 1.3 cm observations of this region, where we detect twelve sources at different evolutionary stages. Most of these sources are only detected at C band with a few sources being detected at both frequency bands (30A, 30E, 30N, 30S and 31). Table \ref{tab:Multiplicity_G35} shows the information about the multiplicity in G35.58, such as the R.A. and Decl. positions, the estimated spectral indices, association with other wavelengths and whether the source is a new radio detection or not. 

\startlongtable
\begin{deluxetable*}{ccccccc}
    \tabletypesize{\footnotesize}
    \tablewidth{0pt}
    \tablecaption{Multiplicity in G35.58-0.03.}
    \label{tab:Multiplicity_G35}
    \tablehead{
    \colhead{Source} & \colhead{R.A (J2000)} & \colhead{Decl. (J2000)} & \colhead{Spectral Index} & \colhead{Association}  & \colhead{New Detection}}
    \startdata
    30A & 18:56:22.52 & +02:20:30.5 & 0.7(0.1) & IR  & Yes\small\textsuperscript{a} \\
    30B & 18:56:22.62 & +02:20:34.7 & $<$-1.1 & \nodata  & Yes\small\textsuperscript{b} \\
    30C & 18:56:22.57 & +02:20:24.3 & $<$-0.3 & \nodata  & Yes\small\textsuperscript{b} \\
    30D & 18:56:22.47 & +02:20:24.8 & $<$-1.2 & \nodata  & Yes\small\textsuperscript{b} \\
    30E & 18:56:22.55 & +02:20:21.4 & -0.5(0.1) & \nodata  & Yes\small\textsuperscript{b} \\
    30F & 18:56:22.43 & +02:20:20.5 & $<$-0.5 & \nodata  & Yes\small\textsuperscript{b} \\
    30G & 18:56:22.39 & +02:20:19.7 & $<$-1.0 & \nodata  & Yes\small\textsuperscript{b} \\
    30H & 18:56:22.46 & +02:20:18.9 & $<$-1.9 & \nodata  & Yes\small\textsuperscript{b} \\
    30J & 18:56:22.19 & +02:20:21.5 & $<$0.6 & \nodata  & Yes\small\textsuperscript{b} \\
    30K & 18:56:22.19 & +02:20:21.5 & $<$0.1 & \nodata  & Yes\small\textsuperscript{b} \\
    30N & 18:56:22.53 & +02:20:27.6 & -0.5(0.4) 
    & IR, mm & Yes\small\textsuperscript{c} \\
    30S & 18:56:22.52 & +02:20:27.2 & 0.2(0.1)
    & IR, mm & Yes\small\textsuperscript{c} \\
    31  & 18:56:22.62 & +02:20:27.9 & 0.0(0.1) & MIR, mm  & No \\
    \enddata
    \tablecomments{ Units of R.A. are hours, minutes, and seconds. Units of Decl. are degrees, arcminutes, and arcseconds. \\
    \textsuperscript{a} These sources have been previously detected at different wavelengths but not in radio continuum. \\
    \textsuperscript{b} These sources have been previously observed but not detected. \\
    \textsuperscript{c} These sources have been previously detected in radio continuum but we are resolving into multiple detections.}
\end{deluxetable*}

\textit{G35.58-0.03 30 and 31:}
Sources 30 and 31 have previously been referred to as UC HII regions \citep[e.g.,][]{1994ApJS...91..659K}.  For source 31, based on its morphology in the centimeter and its estimated spectral index, we agree with its nature being an UC HII region. Moreover, source 31 appears to have already a central protostar based on its association with a bright and compact MIR source (\citetalias{Liu_2019}).

\begin{figure}
\figurenum{3}
\begin{center}
\includegraphics[width=0.9\linewidth]{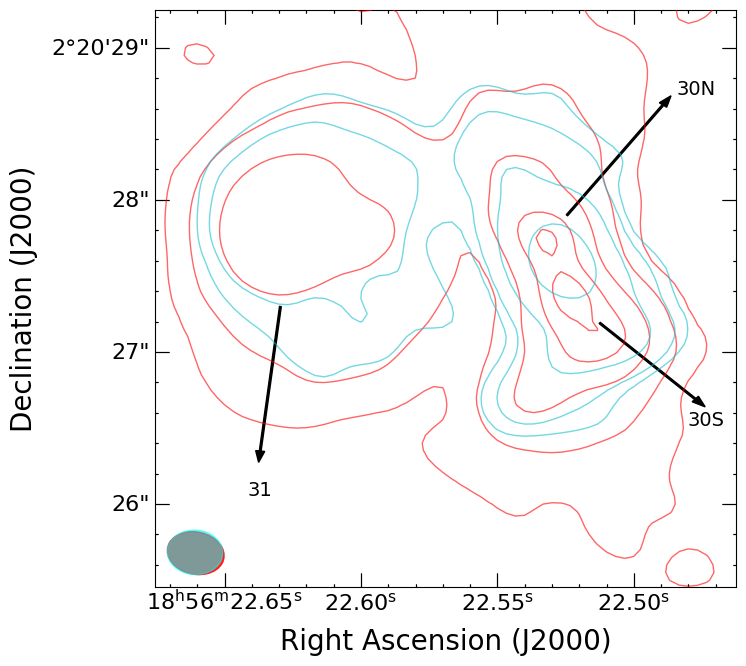}
\caption{VLA continuum maps of the triple system towards G35.58-0.03 observed at 1.3 and 6 cm. The beam sizes are indicated in the lower left-hand corner. The contour levels are at $\sigma \times$(-10, 10, 100, 500, 1200, 1700) for C-band (red) and $\sigma \times$(-10, 10, 25, 100, 500) for K-band (cyan), with the $\sigma$ value reported in Table \ref{tab:SOMA_Sources} \label{fig:G35_Triple}}
\end{center}
\end{figure}

From our observations, source 30 shows a double peak at C-band and a single peak at K-band. At lower frequencies, we marginally resolve it into two components, labeled 30N and 30S (Figure~\ref{fig:G35_Triple}), aligned roughly NE–SW, consistent with the larger-scale molecular outflow previously observed in the region. This together with their association with water masers suggests that 30N and 30S could trace an ionized jet. The estimated spectral index of $\sim$0.7, derived from the inner scale encompassing source 30, is consistent with typical values for ionized jets associated with young stellar objects (YSOs; e.g., \citealt{Reynolds1986Continuum,Anglada1998SpectralSources,Tanaka_2016}). In Table~\ref{tab:Multiplicity_G35}, we also provide estimates of the spectral indices for 30N and 30S. However, higher-resolution observations will be required to confirm the jet scenario. Finally, the association with multiple star formation tracers further suggests that source 30 may be in a younger evolutionary stage than source 31.

\textit{G35.58-0.03 30A:}
Source 30 A appears in the MIR observations as two-pronged diffuse emission in the northern direction (\citetalias{Liu_2019}). We have detected, for the first time, associated centimeter emission with a bubble-like morphology.

\begin{figure*}[ht!]
\figurenum{4}
\begin{center}
\includegraphics[width=0.8\linewidth]{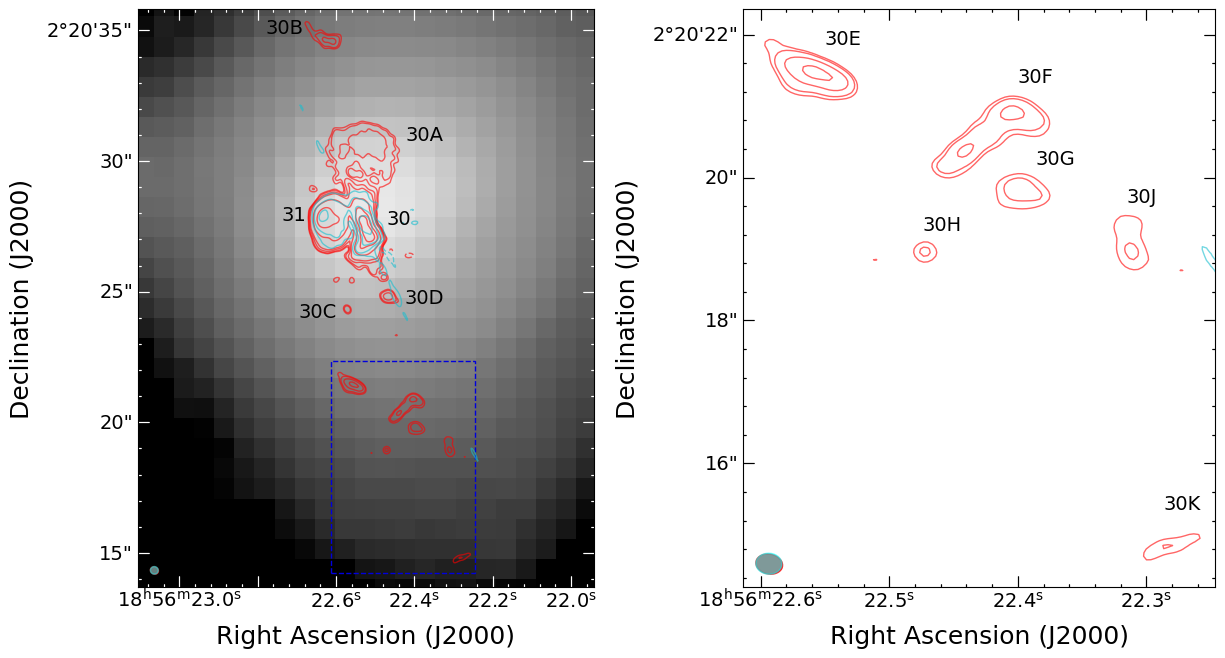}
\caption{VLA continuum 1.3 and 6 cm maps showing the multiplicity of region G35.58-0.03 in the SW of the SOMA scale. The beam sizes are indicated in the lower left-hand corner. The contour levels are at $\sigma \times$(-8, 8, 11, 20, 35) for C-band and $\sigma \times$(-5, 5, 10) for K-band, with the $\sigma$ value reported in Table \ref{tab:SOMA_Sources} \label{fig:G35_SW}}
\end{center}
\end{figure*}

\textit{G35.58-0.03 30B-30K:}
Sources 30B to 30K are presented in Figures \ref{fig:VLA_Contours} and \ref{fig:G35_SW}. To our knowledge, these are new weak radio detections (for more details see Table \ref{tab:Multiplicity_G35}). The majority of the sources are only detected at C band except for 30E that is detected at both C and K-band.
Figure \ref{fig:G35_SW} is centered $\sim$ 2.2$^{\prime\prime}$ to the SW of the SOMA scale (coordinates in Table \ref{tab:SOMA_Sources}).

These new sources present a variety of morphologies: mostly compact sources, but with some of them slightly extended. Upper limit estimates of the spectral indices indicate that all these new sources, except for 30J, are consistent with synchrotron emission. Given their slight alignment with the direction of the molecular outflow, these sources may be part of an ionized jet that is being driven by source 30, which is itself elongated in the same direction.

Despite our suggested classification of an ionized jet for source 30, defining an intermediate scale for this source is not feasible. The extent of the presumed jet is already enclosed within the SOMA scale, with the caveat that this scale contains the UCHII region in source 31 and the bubble-like morphology source 30A. Any intermediate scale would face the same issue; therefore, we omitted this scale for this particular case.

The high multiplicity observed in this region raises the question of whether these detections are real or artifacts. In general, we expect artifacts to exhibit symmetric properties, but, as shown in Figures \ref{fig:VLA_Contours}, \ref{fig:G35_Triple}, and \ref{fig:G35_SW}, we can see that this is not the case for most of our detections. Furthermore, based on estimates of calibration and deconvolution errors, we find that Source 30B is at least 25 times brighter than what would be expected for a Point Spread Functions (PSF) artifact. This implies that detections at least 10 times fainter than Source 30B could plausibly be artifacts. Among our sources, this criterion only applies to Source 30J, which is approximately 10 times fainter than 30B. Therefore, we conclude that most of our detections are likely real.

\subsubsection{IRAS 16562-3959}

IRAS 16562-3959, also known as G345.49+1.47, is located at a distance of 1.7 kpc and hosts a massive core containing a massive star in an early stage of evolution \citep{2010ApJ...725..734G}. \cite{2011ApJ...736..150G} reported a quadrupolar molecular outflow traced by CO. This morphology is explained by the presence of two collimated bipolar outflows, one lying in the SE–NW direction related to the ionized jet detected by \cite{2010ApJ...725..734G}, and another in the N–S direction that could be related to the unresolved millimeter source 13 in \cite{2014ApJ...796..117G}.

For our analysis we used A. Guzman's ATCA observations at 5.5, 9, 17, and 19 GHz from \cite{2016ApJ...826..208G}, where they reported nine radio detections associated with this region. The central source has a counterpart at 3~mm and X-ray (source 10 at 3~mm and source 161 in X-ray) \citep{2014ApJ...796..117G, 2020ApJ...888..118M}, and has also been detected in the MIR by the \citetalias{Liu_2019} study.

We focus on the regions IRAS 16562-3959 and IRAS 16562-3959 N. The latter was added to the sample in \citetalias{Liu_2020} due to the detection of a point source in the IR maps, based on the observations analyzed in \citetalias{Liu_2019}.

\textit{IRAS 16562-3959:} For the purpose of this study, we only focus on the sources within the SOMA scale (see Figure \ref{fig:VLA_Contours}). Detections C, I-E, and I-W all exhibit compact morphologies. As described by \citet{2014ApJ...796..117G}, this region shows characteristics consistent with an ionized radio jet, driven by the central free–free emission source (C) and accompanied by two lobes (I-E and I-W) of nonthermal emission. Given the clear jet-like morphology we also measure the fluxes using an intermediate scale, as previously described, that encompass the central source (detection C) and the two inner radio lobes.

Sources 1, 2, and 3 exhibit compact morphologies. \citet{2016ApJ...826..208G} discuss the nature of these detections in Section 4.2.2 of their paper. In summary, Source 1 may be extragalactic, Source 2 is likely an HC H II region associated with a young high-mass star, and the free–free emission from Source 3 originates from a low-mass YSO.

Based on our criteria, and for completeness, we report all detections above a $5\sigma$ threshold, adopting the same signal-to-noise definition as \citet{2016ApJ...826..208G}. As a result, we identify a new source, source 4 (following the convention of \citealt{2016ApJ...826..208G}), which is detected only at 5 and 9 GHz, shows a slight elongation to the south (Figure~\ref{fig:VLA_Contours}), and was not previously reported. We estimate an upper limit to its spectral index of $\alpha < 1.3$.

\textit{IRAS 16562-3959 N:} In these observations we were not able to detect source IRAS 16562-3959 N which is a MIR source reported by \citetalias{Liu_2020} as a secondary region around IRAS 16562-3959. This point source was detected using IR data and, moreover, we see this new detection at the shortest wavelengths (MIR), but not in the longest wavelengths (FIR) of the observations analyzed in \citetalias{Liu_2019}. 
  
\subsubsection{G305.20+0.21}

The G305.2+0.2 complex is a vast high-mass star-forming region located at a distance of 4.1 kpc \citep{2017MNRAS.465.1095K} in the southern Galactic Plane. Within the region, there are two locations separated by 22$^{\prime\prime}$ named G305A (G305.21+0.21) and G305B (G305.20+0.21) where Class II methanol (\ch{CH3OH}) masers have been detected \citep{1993ApJ...412..222N,2006MNRAS.365..321W, 2007MNRAS.380.1703W}. Additionally, G305C has been recently reported by \citetalias{Liu_2019} as an IR source located at 14$^{\prime\prime}$ to the east of G305B, and it is detected at all SOFIA wavelengths showing NIR counterparts, as well.

From previous radio continuum observations toward G305.20$+$0.21 there are reports of two detections, which are prominent extended HII regions labeled as G305HII(SE) and G305HII by \cite{2007MNRAS.380.1703W}. These regions are located about 30$^{\prime\prime}$ to the southeast of G305A and 15$^{\prime\prime}$ to the southwest of G305B, respectively \citep{2007MNRAS.380.1703W, 1998MNRAS.300.1131P}.

Figure \ref{fig:VLA_Contours} show the observations toward this region, where the centimeter data images were made using a minimum baseline length of 1.5~km.  
The main reason for doing this was to improve the detections of the compact radio emission associated with G305A and G305B, which are our sources of interest, by filtering out the bright, extended emission from the two HII regions (i.e., G305HII(SE) and G305HII) located to the south of each of them.
  
\textit{G305A (G305.21+0.21):} This source is prominent at  MIR and FIR \citepalias{Liu_2019} and has been observed at 8.6 and 18~GHz, but no radio detections have been reported at sensitivity levels of 0.9~mJy and 0.22~mJy, respectively \citep{1998MNRAS.300.1131P,2007MNRAS.380.1703W}. With our ATCA observations at 5.5, 9, 17 and 22.8 GHz, we report a single detection within the region G305A that we label as G305A ATCA-1 (see Figure \ref{fig:VLA_Contours}: G305.21+0.21 (G305A)). This source coincides with the methanol maser reported by \cite{2006MNRAS.365..321W} and shows a slight elongation at the lower frequencies. \cite{2021A&A...645A.110Y, 2022A&A...658A.160Y} reported CO molecular outflows towards this region in the ATLASGAL clump scale.

For G305A we report an estimated spectral index of $\alpha\sim$ 0.3, a result that indicates thermal emission. \citetalias{Liu_2019} suggested that G305A is a much younger and more embedded source than G305B and in a hot core phase, rich in molecular tracers \citep{2007MNRAS.380.1703W}. Considering the spectral index value and star-formation tracers such as reported masers and molecular outflows, we suggest that this source is most likely a radio jet.

\textit{G305B (G305.20+0.21):} Region G305B is the brightest MIR source and it is associated with  methanol masers. However, no radio emission has been previously detected towards this region \citep{2007MNRAS.380.1703W}. In our ATCA observations (see Figure \ref{fig:VLA_Contours}) we report 4 sources labeled  G305B ATCA-1, G305B ATCA-2, G305B ATCA-3 and G305B ATCA-4 (hereafter sources 1, 2, 3 and 4).
In Table \ref{tab:Multiplicity_G305} we present more information about the multiplicity in G305B such as the R.A. and Decl. positions, the estimated spectral index, association with other wavelengths, and whether this source is a new radio detection or not.

\startlongtable
\begin{deluxetable*}{ccccccc}
    \tabletypesize{\footnotesize}
    \tablewidth{0pt}
    \tablecaption{Multiplicity in G305.20+0.21 (G305B)}
    \label{tab:Multiplicity_G305}
    \tablehead{
    \colhead{Source} & \colhead{R.A (J2000)} & \colhead{Decl. (J2000)} & \colhead{Spectral Index} & \colhead{Association}  & \colhead{New Detection}}
    \startdata
    G305B ATCA-1N   & 13:11:10.58 & $-$62:34:37.8 & $<$0.4 & IR & Yes\small\textsuperscript{a} \\
    G305B ATCA-1S   & 13:11:10.56 & $-$62:34:38.8 & $<$-1.0 & IR & Yes\small\textsuperscript{a} \\
    G305B ATCA-2    & 13:11:10.04 & $-$62:34:40.1 & -0.8(0.2) &  \nodata & Yes\small\textsuperscript{b} \\
    G305B ATCA-3    & 13:11:09.63 & $-$62:34:38.7 & -0.6(0.2) & \nodata & Yes\small\textsuperscript{b} \\
    G305B ATCA-4    & 13:11:10.99 & $-$62:34:37.8 & $<$-1.7 & \nodata & Yes\small\textsuperscript{b} \\
    \enddata
    \tablecomments{Units of R.A. are hours, minutes, and seconds. Units of Decl. are degrees, arcminutes, and arcseconds. \\
    \textsuperscript{a} These sources have been previously detected at different wavelengths but not in radio continuum. \\
    \textsuperscript{b} These sources have been observed but not detected.}
\end{deluxetable*}

Source 1 is the main detection in the region, and in our ATCA observations, this source is seen at all frequencies (5.5, 9, 17 and 22.8~GHz) and exhibits a morphology with a slight elongation in the NE-SW direction. \citetalias{Liu_2019} was able to resolve this source into two sources (G305B1 and G305B2) using high-spatial-resolution \textit{Gemini} observations, finding morphologies that are also aligned in the NE-SW direction. They suggested that both sources are part of an outflow cavity. In our ATCA observations at the lowest frequencies (5.5 and 9~GHz), source 1 has two peaks, 1N and 1S (see inset in Figure \ref{fig:VLA_Contours}), that appear to be associated with their MIR counterpart. Our global spectral index estimate of Source 1 is dominated by non-thermal emission ($\alpha =$ -0.7), suggesting the existence of a high-speed jet in the region. Given the orientation of the sources, the driving source of this jet is likely source 1N, which is the brightest source, and we estimate a rough spectral index of $\alpha \sim0.4$.

To our knowledge, sources 2, 3 and 4 are also new detections in the radio continuum. Source 2 is seen in all our ATCA observations and, similarly to source 1, shows a slight elongation in the NE direction and exhibits a spectral index that indicates non-thermal emission.

Source 4 is a faint compact eastern source only seen at the lowest frequencies (5.5 and 9~GHz), and we report a limit on its spectral index (see Table \ref{tab:Multiplicity_G305}). On the other hand, source 3 is seen at all frequencies and presents an elongated morphology in the SE-NW  direction that appears as a double peak at the longest contour levels shown in Figure \ref{fig:VLA_Contours}. This source may be part of the HII region in \cite{1999MNRAS.309..905W}, but in our observations we report it as an independent source.
We report a negative spectral index ($\alpha \sim$ -0.6) suggesting non-thermal emission.

Since Sources 2 and 4 are aligned in the same SW-NW direction, exhibit relatively compact morphologies and negative spectral indices, we conclude that the most likely nature of these sources is that they are non-thermal knots, centered in Source 1N. Following our convention for sources with jet-like morphology, we measured the flux densities at an intermediate scale that encompasses the central source and the jet knots (Sources 1, 2, and 4). At this scale, we estimate an upper limit on the spectral index of $\alpha < -1.2$, which is indicative of a non-thermal jet in the region.

Additionally, source G305C is within the SOMA scale for region G305B, but we are not detecting any radio source at our sensitivities levels. It is important to highlight that due to the minimum baseline length used in our images toward these regions, the flux measured for the SOMA scale corresponds to a lower limit, specially for source G305B due to the proximity to the HII region which has been filtered out in our images.

\subsubsection{G49.27-0.34}

G49.27-0.34 is a protostar located in an Infrared Dark Cloud (IRDC) at a kinematic distance of 5.55~kpc \citep{2009ApJ...702.1615C}. To our knowledge, no molecular outflow has been detected toward this region to date. Two Class I 44 GHz \ch{CH3OH} masers spots have been detected in this region while no 6.7 GHz \ch{CH3OH}, 25 GHz \ch{CH3OH} or thermal masers have been detected \citep{2009ApJ...702.1615C, 2017ApJS..230...22T}.

In our 1.3 and 6 cm observations we report 5 detections within the SOMA scale. Table \ref{tab:Multiplicity_G49} shows some information about the multiplicity in G49.27-0.34, i.e., the coordinates, estimate spectral index, association at other wavelengths, and whether or not a source is a new radio detection. We discuss each detection below.

\startlongtable
\begin{deluxetable*}{cccccc}
    \tabletypesize{\footnotesize}
    \tablewidth{0pt}
    \tablecaption{Multiplicity in G49.27-0.34.}
    \label{tab:Multiplicity_G49}
    \tablehead{
    \colhead{Source} & \colhead{R.A (J2000)} & \colhead{Decl. (J2000)} & \colhead{Spectral Index} & \colhead{Association} & \colhead{New Detection}}
    \startdata
    CM1  & 19:23:06.87 & +14:20:17.3 & 0.8(0.1) & IR & No \\
    CM2A & 19:23:06.66 & +14:20:11.6 & 1.6(0.2) & IR & Yes\small\textsuperscript{a} \\
    CM2B & 19:23:06.62 & +14:20:12.1 & $>$ 0.7 & IR & Yes\small\textsuperscript{a} \\
    CM2C & 19:23:06.58 & +14:20:11.9 & 0.0(0.1) & IR & Yes\small\textsuperscript{a} \\
    CM3  & 19:23:06.71 & +14:19:59.0 & $<$ 0.5  & IR & Yes\small\textsuperscript{b} \\
    \enddata
    \tablecomments{ Units of R.A. are hours, minutes, and seconds. Units of Decl. are degrees, arcminutes, and arcseconds. \\
    \textsuperscript{a} These sources have been previously detected in radio continuum but we are resolving into multiple detections.  \\
    \textsuperscript{b} These sources have been previously detected at different wavelengths but not in radio continuum.}
\end{deluxetable*}

\textit{CM1:} Source CM1 is an extended source  
previously seen at 1.3, 3.6 and 20~cm \citep{2017ApJS..230...22T, 2011ApJ...743...56C, 1994ApJS...91..713M} and it is coincident with MIPS 24~$\mu$m emission \citep{2009PASP..121...76C}. 
There are no reports of association with
Class I 44 GHz \ch{CH3OH} masers.

\cite{2011ApJ...743...56C} suggests that CM1 is a single ionizing star of spectral type B0V based on optically thin free–free emission and the ionizing photon flux. Our 1.3~cm flux density values are one order of magnitude less than that reported by \cite{2011ApJ...743...56C, 2017ApJS..230...22T} with VLA observations and resolutions of $\sim0.8$$^{\prime\prime}$ (2.63 and 1.62 mJy). This difference is likely due to the different resolutions. 

\textit{CM2:} The MIR peak observed in \citetalias{Liu_2019} is associated with source CM2, previously reported as an unresolved,  4$\sigma$ detection at 0.027 mJy/beam rms at 3.6 cm \citep{2011ApJ...743...56C}.  Additionally, \cite{2011ApJ...743...56C} estimated a spectral index limit of $<$0.2 at a resolution of $\sim0.9$$^{\prime\prime}$.
Using our VLA data, we are able to resolve source CM2 into three components, which we have labeled as CM2A, CM2B and CM2C, with all sources detected at both 1.3 and 6~cm observations.

CM2A is the central and brightest source at 1.3~cm among the components of CM2. It shows a compact morphology and a very steep spectral index of 1.6.

CM2B is detected as an unresolved source, located $\sim$ 0.7$^{\prime\prime}$ SW of the central component (CM2A) and it is the weakest detection in this system. We estimated a lower limit for the spectral index of 0.7.  

CM2C is located about 1.2$^{\prime\prime}$ of the central component in the same direction as CM2B. It presents a compact morphology with a flat spectral index suggesting thin free-free emission. 

Given the extended morphology of the CM2A-CM2C sources at low frequencies, we also measure the flux densities using an intermediate scale that incorporates the three components. The estimated spectral index at this scale yields a value of 0.7, given an indication that CM2 could be a radio jet with CM2A being the driven source of this jet.

\textit{CM3:} From the contour maps in \cite{2011ApJ...743...56C, 2017ApJS..230...22T}, source CM3 shows MIPS 24~$\mu$m emission and \citetalias{Liu_2019} also detected this source in the SOFIA IR maps, but not in the NIR. This source has no previous radio detections at 1.3 and 3.6~cm. In our observations we are able to detect this source, but only at 6~cm (C-band).

CM3 exhibits a compact morphology and we report a relative flat spectral index with an upper limit of 0.5, indicative of thermal emission. To our knowledge, we are detecting this source for the first time in radio frequencies and its naming follows the convention of the main region, G49.27-0.34. This source may be a YSO at a more evolved stage than CM2, although additional observations are required for a definitive assessment.

\subsubsection{G339.88-1.26}
 
G339.88-1.26 (or IRAS 16484-4603) is a massive protostar located at 2.1~kpc \citep{2015ApJ...805..129K}, that has been previously observed at a variety of wavelengths. \cite{2002ApJ...564..327D} resolved the central MIR emission into three peaks (1A, 1B and 1C) and radio continuum observations have identified the central source is associated with a radio jet and the other two sources are associated with radio outflow lobes \citep{1996MNRAS.279..101E, 2016MNRAS.460.1039P}, label as C, NE and SW. Using ALMA data, \cite{2019ApJ...873...73Z} reported two molecular outflows towards this region positioned at the source C. One of this outflows is located in the northeast-southwest orientation and exhibits the same angle as the ionized outflow reported by \cite{2016MNRAS.460.1039P}.

With 19.7 $\mu$m SOFIA data, both molecular outflows are detected; however, at longer wavelengths, only the molecular outflow associated with the ionized outflow is observed. \citetalias{Liu_2019} suggests this could be due to extinction, and that the detection of red and blue shifted emission on both sides suggests a near side-on view of the outflows.

For our analysis, we used the ATCA observations at 5.5, 9.0, 17.0, and 22.8 GHz reported by \citet{2016MNRAS.460.1039P}. We focus on the three sources within the SOMA scale i.e., NE, C, and SW, identified by \citet{2016MNRAS.460.1039P}. Source C appears compact at higher frequencies but shows a slight extension at lower frequencies, particularly at 5.5 GHz. In contrast, Sources NE and SW exhibit relatively compact morphologies, with Source SW being especially compact.

As mentioned by \cite{2016MNRAS.460.1039P}, source G339.88-1.26 shows characteristics consistent with an ionized radio jet driven by the central source C and two lobes of non-thermal emission. Given this jet-like morphology of the region we also measure the flux densities using an intermediate scale that encompass the three sources. We report a spectral index of $\alpha \sim-$ 0.1 for the intermediate scale suggesting that the non-thermal emission from the lobes is more prevalent at this scale.

\subsection{Radio SEDs}\label{Radio_SEDs}

\begin{figure*}[ht!]
\figurenum{5}
\begin{center}
\includegraphics[width=0.32\linewidth]{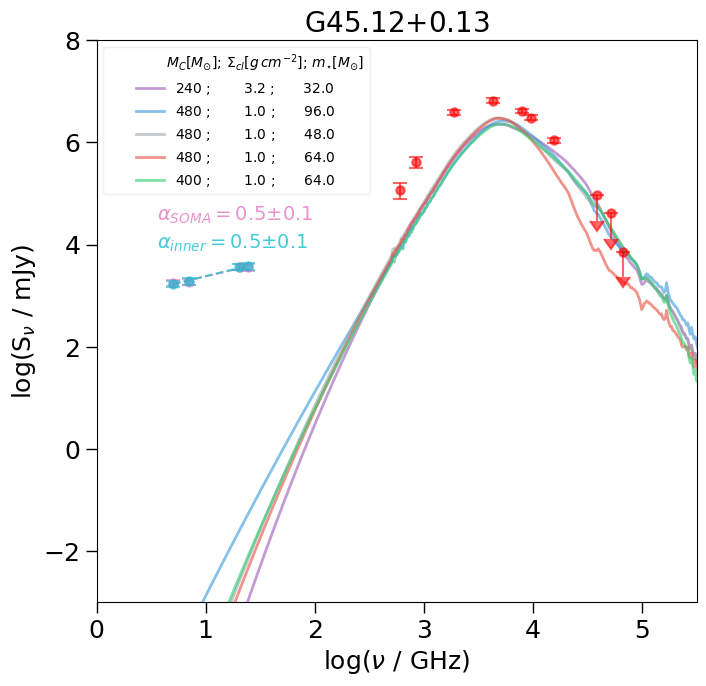}\quad\includegraphics[width=0.32\linewidth]{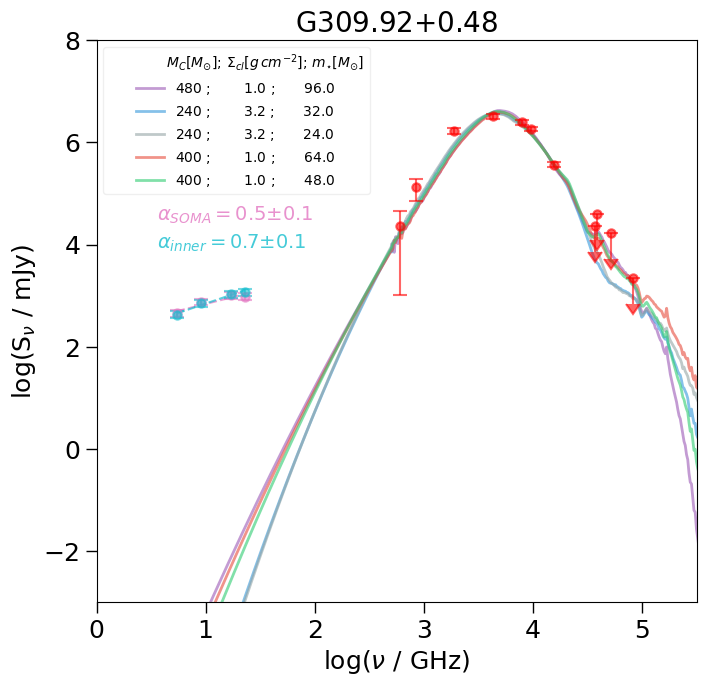}\quad\includegraphics[width=0.32\linewidth]{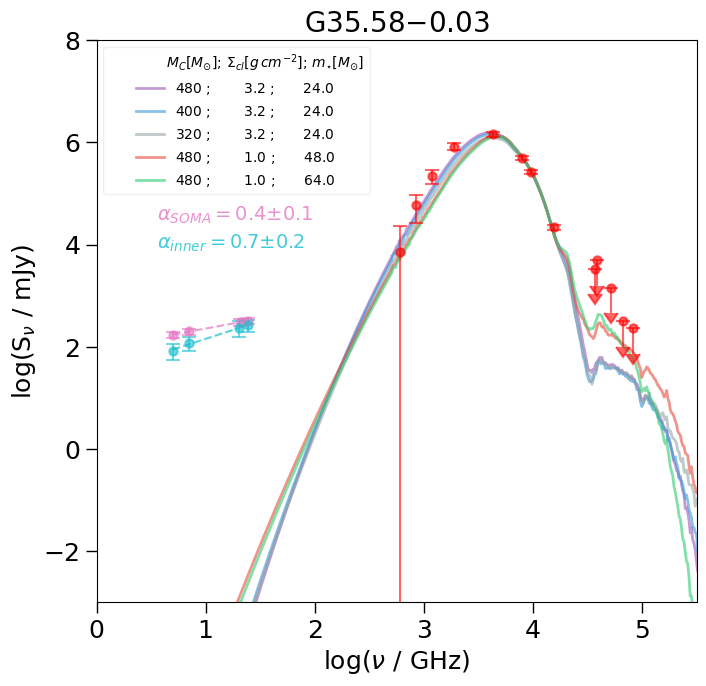}\\
\includegraphics[width=0.32\linewidth]{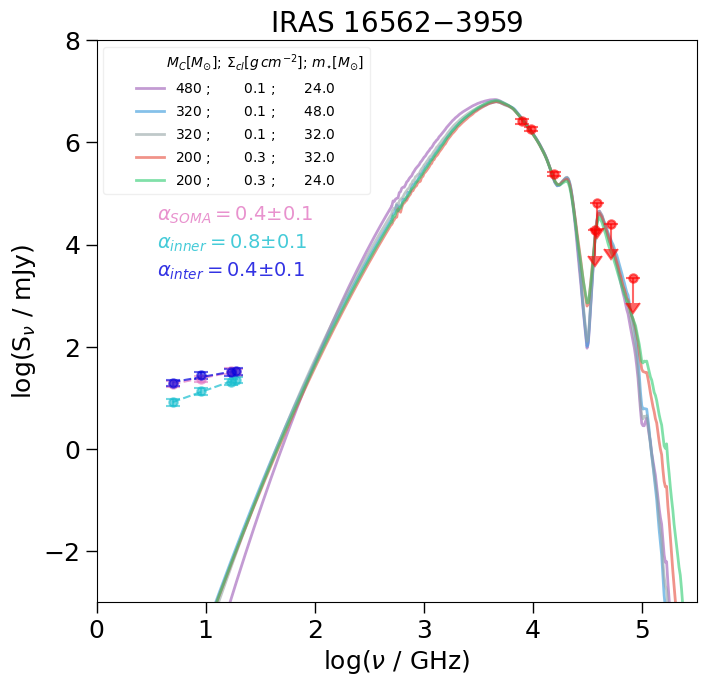}\quad\includegraphics[width=0.32\linewidth]{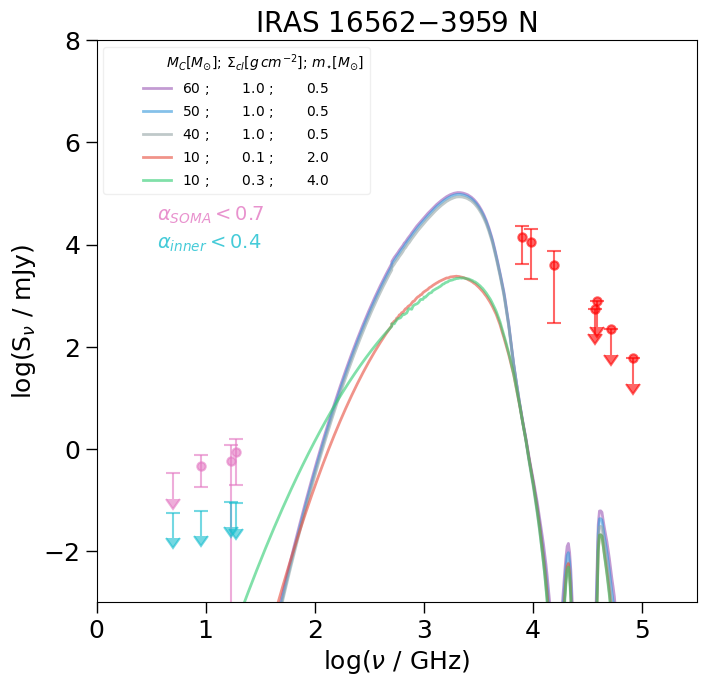}\quad\includegraphics[width=0.32\linewidth]{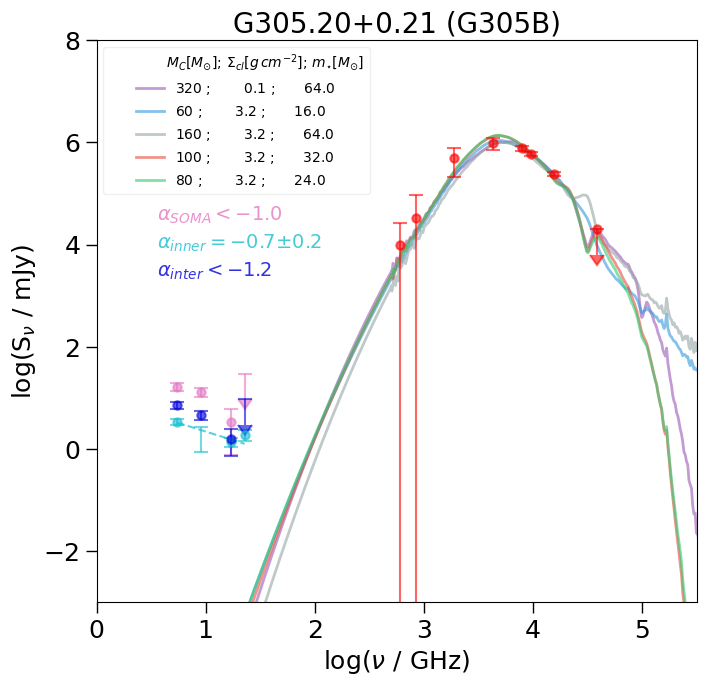}\\
\includegraphics[width=0.32\linewidth]{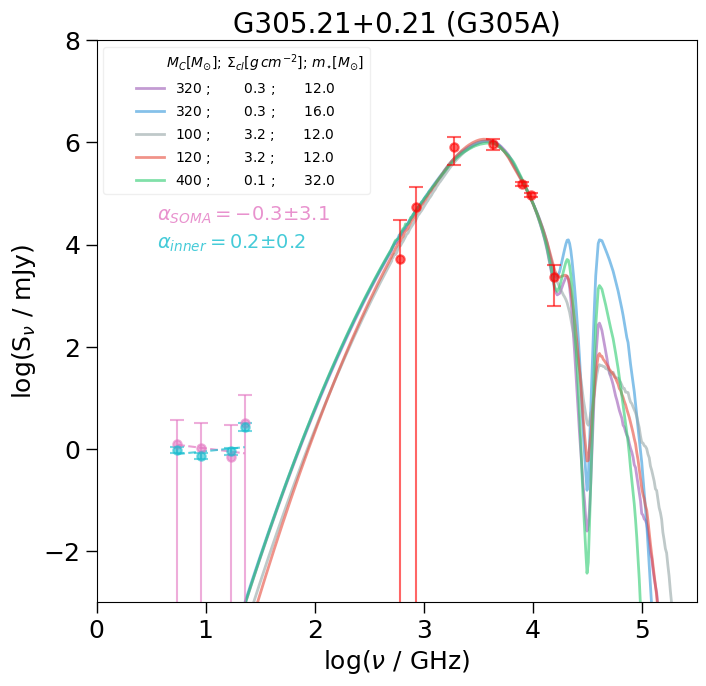}\quad\includegraphics[width=0.32\linewidth]{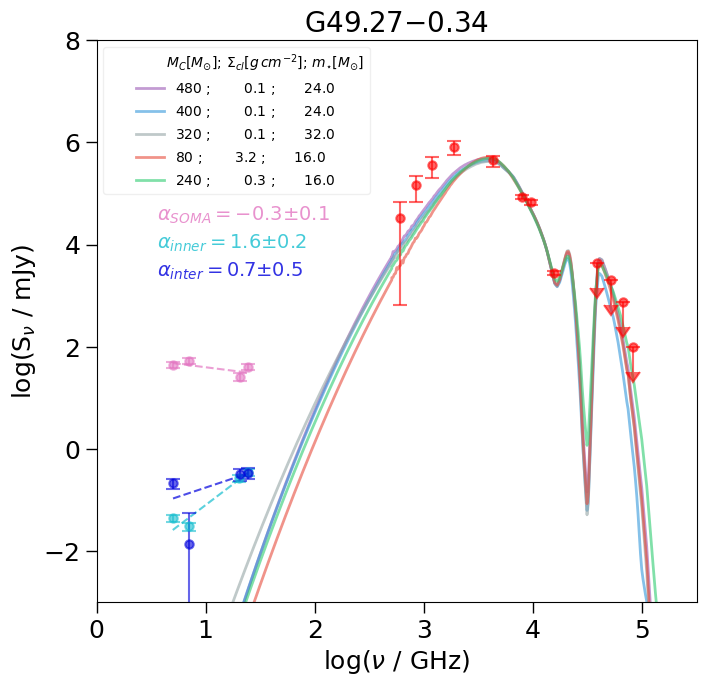}\quad\includegraphics[width=0.32\linewidth]{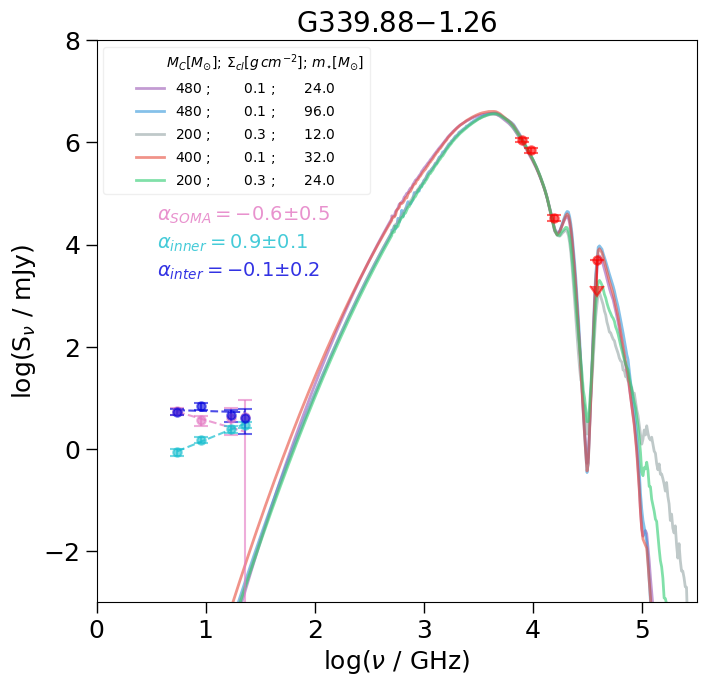}\\ 
\caption{Extended spectral energy distributions (E-SEDs) of SOMA protostars, consisting of radio and IR SEDs. Red circles show IR data for the SOMA apertures as measured by \citetalias{Liu_2019}. The other colored circles correspond to the centimeter flux density as a function of the frequency at each scale (magenta: \textit{SOMA}; blue: \textit{Intermediate}; cyan: \textit{Inner}). Error bars are explained in \S\ref{Radio_SEDs}. The solid lines show the five best fits (see legend) to the IR data SED from the \cite{Zhang_2018} models as fit by \citetalias{Fedriani_2023} (see Appendix in \citetalias{Fedriani_2023}), and the dashed lines are the best fit of the radio data using a power law of the form $S_{\nu} \propto \nu^{\alpha}$. \label{fig:Radio_SEDs}}
\end{center}
\end{figure*}

Building upon the results of \citetalias{Liu_2019}, in Figure \ref{fig:Radio_SEDs} we present the E-SEDs, i.e., radio + IR SEDs, for our nine protostellar sources. The dashed lines correspond to the best fit to the radio continuum data using a power law of the form $S_{\nu} \propto \nu^{\alpha}$, where $\alpha$ is the spectral index at the various scales: \textit{"SOMA"}; \textit{“Intermediate”}; and \textit{“Inner”}, as described in section \ref{subsec:results}. The spectral index was calculated using the flux density at the central frequencies from the images, so $\alpha$ is calculated over a wide frequency range ($>$20~GHz). The uncertainty in the spectral index was calculated with a Monte Carlo simulation that bootstrapped the flux density uncertainties. We estimated an upper limit in the spectral index for non-detections at higher frequencies using a value of $S_{\nu}$ of 3$\sigma$. 

It is important to mention that at higher frequencies the free-free emission is likely contaminated by dust emission (see \cite{2016ApJ...832..187B}), therefore adding some additional systematic uncertainty to the measurement of the free-free fluxes. 
 
\section{Analysis and Discussion}\label{analysis}

The main goal of this paper is to analyze the morphology, multiplicity, and the estimated spectral indices of the sources of the SOMA II sample to better understand the nature of the sources using radio observations at different frequencies. Our sample consists of regions located at distances greater than 1~kpc, with some exhibiting low radio flux densities ($<$3 mJy), such as G305.20+0.21 (G305B), G305.21+0.21 (G305A), G49.27-0.34, and G339.88-1.26. In contrast to \citetalias{Rosero_2019}, seven of the eight sources analyzed showed relatively low radio flux densities.  

With the available data, we aim to assess the nature of the sources based on their morphology and spectral index. For sources that have been previously analyzed, we refer to past studies that have characterized their nature and compare our results, specifically in terms of morphology and spectral index (see \S\ref{Morph}). This comparison allows us to refine our understanding of the radio emission mechanisms associated with each source.

\subsection{Multiplicity and new detections}

Most of the sources analyzed in this paper present a high level of multiplicity, except for G305A which only has one detection within the SOMA scale. Furthermore, no source has been detected in IRAS 16562 N at the sensitivity level ($\sim$21 $\mu$Jy -- 28 $\mu$Jy) of the analyzed ATCA data.

Across our 9 target regions, we detected a total of 37 sources. See \S\ref{Morph} for individual analysis of each source. About 38\% of the sources are new detections, hence in \S\ref{Morph} we made the notation of two types of new detections: (A) sources that have been previously observed and detected at different wavelengths, but not at radio frequencies; and (B) sources that have never been detected at any wavelength. This type of new detection could be regions that have been observed at radio but remained undetected due to limitations of sensitivity or possible variable sources. Detections of regions in which we present the first radio observations also fall in this category.

Of the 37 sources, we report 13 new detections (9 in G35.58-0.03, 1 in IRAS 16562-3959, and 3 in G305B or G305.20+0.2) classified as Type B detections, which, to the best of our knowledge, have never been detected at any wavelength.

To make sure we are not missing any UCHII regions, we estimate the physical properties from the 5~GHz sensitivity limit (rms) using equation 1 and 3 from \cite{1994ApJS...91..659K} and equation A.2.3 from \cite{Panagia1978}. These equations assume spherical symmetry and optically thin emission from a uniform-density plasma with \( T_e = 10^4\,\mathrm{K} \). The results are listed in Table \ref{tab:ZAMs_Calc}, where column 1 is the region name, column 2 is the logarithm of the Lyman continuum flux (\( \log N'_{\mathrm{Ly}} \)) required for ionization. With the \( \log N'_{\mathrm{Ly}} \) and the results from Table II in \cite{1973AJ.....78..929P}, we estimate the spectral type of the ionizing star (listed in column 3), further assuming that a single ZAMS star is photoionizing the nebula and producing the Lyman continuum flux. The distances used for these calculations are listed in Table \ref{tab:SOMA_Sources}. Column 4 is the estimate mass of the spectral type of the ionizing star from \cite{2013ApJS..208....9P}.

\begin{deluxetable}{cccc}
    \tabletypesize{\footnotesize}
    \tablewidth{0pt}
    \tablecaption{Parameters from Radio Continuum at the sensitivity limit at 5 GHz. \label{tab:ZAMs_Calc}}
    \tablehead{
    \colhead{Region} & \colhead{$\mathrm{logN'_{Ly}}$\small\textsuperscript{a}} & \colhead{Sp\small\textsuperscript{b}} & \colhead{$\mathrm{M_{\odot}}$\small\textsuperscript{c}}}
    \startdata
    G45.12+0.13         &  45.34 & B1   & 11.8 \\
    G309.92+0.48        &  46.03 & B0.5 & 14.8 \\ 
    G35.58-0.03         &  45.49 & B1   & 11.8 \\ 
    IRAS 16562-3959     &  43.38 & B3   & 5.4 \\ 
    IRAS 16562-3959 N   &  43.40 & B3   & 5.4 \\ 
    G305.20+0.21        &  44.98 & B1   & 11.8 \\ 
    G305.20+0.21 A      &  44.96 & B1   & 11.8 \\ 
    G49.27-0.34         &  44.07 & B3   & 5.4 \\ 
    G339.88-1.26        &  44.01 & B3   & 5.4 \\ 
    \enddata
    \tablecomments{\\
    \textsuperscript{a} Values calculated using equations from \cite{1994ApJS...91..659K, Panagia1978}.\\
    \textsuperscript{b} Spectral type estimation made from table II in \cite{1973AJ.....78..929P}.\\
    \textsuperscript{c} The estimate of the mass was taken from \cite{2013ApJS..208....9P}}
\end{deluxetable}

The results from Table \ref{tab:ZAMs_Calc} give us an estimation of the lowest mass ZAMS star that we can detect with our current observations. This serves as completeness parameter on the multiplicity at the different regions, especially for region IRAS 16562-3959 N where we do not report any detections and in this case the completeness level goes down to around 5.4~$M_\odot$.

\subsection{Nature of the sources}

By using previous studies on these sources, as well as results from our radio observations, we have assessed the nature of each individual source (see \S\ref{Morph}). From a sample point of view of the SOMA II sources, out of the 37 detections it is important to mention that three of them are well defined radio jets given their morphology and the estimated spectral index. For example, IRAS 16562-3959 and G339.88-1.26, are both very well previously studied ionized jets. Both have a central source that exhibits a jet-like morphology and a spectral index in the range for typical ionized jets associated with YSOs \citep{Reynolds1986Continuum, Anglada1998SpectralSources, Tanaka_2016}, along with two additional sources classified as lobes that display non-thermal emission due to the synchrotron radiation. For G49.27-0.34 CM2A we report a spectral index in the typical range of ionized jets and two additional sources relatively close to the driving jet (CM2B at 0.7$^{\prime\prime}$ and CM2C at 1.2$^{\prime\prime}$), but both companions exhibit thermal emission.

We also propose a radio jet scenario for sources G35.58–0.03, G305.21+0.21 (G305A) and G305.20+0.21 (G305B), based on the calculated spectral index, observed extended morphology, association with extended molecular outflows and in some cases evidence of knots from a non-thermal jet in the direction of the elongation. However, this classification remains less certain compared to other radio jets in the sample, and additional observations are needed to fully resolve and confirm the presence of the proposed non-thermal jets.

Additionally, we have three UC HII regions (G45.12+0.13 S14, G35.58-0.03 31 and IRAS 16562-3959 2), following the suggestion of previous authors and the morphology of our detections and the reported spectral indices. The remaining sources are point sources being companion sources from the central detection or sources of an unknown nature, that could be of extragalactic nature or lower-mass YSOs. Most of them display a flat or negative spectral index suggesting non-thermal emission. Table \ref{tab:nature} summarizes all the sources reported in this paper along with their suggested nature. Column 1 lists the region, Column 2 the corresponding source name, and Column 3 the suggested nature of the source.

\begin{deluxetable}{ccc}
    \tabletypesize{\footnotesize}
    \tablewidth{0pt}
    \tablecaption{Summary of the nature of the sources reported in the SOMA II sample. \label{tab:nature}}
    \tablehead{\colhead{Region} & \colhead{Source} & \colhead{Nature}}
    \startdata
    G45.12+0.13          & S14      & UC HII Region\\
    G45.12+0.13          & S20      & Extragalactic \\ 
    G309.92+0.48         & ATCA1    & Optically Thick emitter \\
    G309.92+0.48         & ATCA2    & Undetermined \\ 
    G35.58-0.03          & 30A      & Undetermined  \\ 
    G35.58-0.03          & 30B      & Jet Knot (Candidate) \\ 
    G35.58-0.03          & 30C      & Jet Knot (Candidate) \\ 
    G35.58-0.03          & 30D      & Jet Knot (Candidate) \\ 
    G35.58-0.03          & 30E      & Jet Knot (Candidate) \\ 
    G35.58-0.03          & 30F      & Jet Knot (Candidate) \\ 
    G35.58-0.03          & 30G      & Jet Knot (Candidate) \\ 
    G35.58-0.03          & 30H      & Jet Knot (Candidate) \\ 
    G35.58-0.03          & 30J      & Jet Knot (Candidate) \\ 
    G35.58-0.03          & 30K      & Jet Knot (Candidate) \\ 
    G35.58-0.03          & 30N      & Unresolved Jet Candidate \\ 
    G35.58-0.03          & 30S      & Unresolved Jet Candidate \\ 
    G35.58-0.03          & 31       & UC HII Region\\
    IRAS 16562-3959      & C        & Ionized Radio Jet \\
    IRAS 16562-3959      & I-E      & Non-thermal Lobe \\
    IRAS 16562-3959      & I-W      & Non-thermal Lobe \\
    IRAS 16562-3959      & 1        & Extragalactic \\
    IRAS 16562-3959      & 2        & HC HII Region \\
    IRAS 16562-3959      & 3        & Low-mass YSO \\
    IRAS 16562-3959      & 4        & Undetermined  \\
    G305.21+0.21 (G305A) & ATCA-1   & Jet Candidate \\
    G305.20+0.21 (G305B) & ATCA-1   & Jet Candidate \\
    G305.20+0.21 (G305B) & ATCA-2   & Jet Knot (Candidate) \\
    G305.20+0.21 (G305B) & ATCA-3   & Part of the HII Region \\
    G305.20+0.21 (G305B) & ATCA-4   & Jet Knot (Candidate) \\
    G49.27-0.34          & CM1      & B0V Ionizing Star \\
    G49.27-0.34          & CM2A     & Jet Candidate \\
    G49.27-0.34          & CM2B     & Jet Knot (Candidate) \\
    G49.27-0.34          & CM2C     & Jet Knot (Candidate) \\
    G49.27-0.34          & CM3      & Undetermined \\
    G339.88-1.26         & C        & Ionized Radio Jet \\
    G339.88-1.26         & NE       & Non-thermal Lobe \\
    G339.88-1.26         & SW       & Non-thermal Lobe \\
    \enddata
\end{deluxetable}

In terms of multiplicity, we find results similar to \citetalias{Rosero_2019}, where the analyzed regions exhibit high multiplicity, with multiple sources detected within the SOMA scale. In \citetalias{Rosero_2019}, seven out of eight regions showed at least two detections. In our sample, seven out of nine regions also show more than one detection, with G305.21+0.21 (G305A) having a single detection and IRAS 16562-3959 N showing no detections, likely due to our sensitivity limit.

In \citetalias{Rosero_2019}, we reported different evolutionary stages and nature of the sample, which included radio jets, HII regions, and variable stars. Four of the seven sources in that initial sample exhibited characteristics consistent with radio jets, such as elongation in the same direction as the associated molecular outflow with a central ionized jet and knots in aligned detections. In the sample presented in this paper, we observe a similar distribution: six of the nine regions show evidence of hosting a radio jet on the SOMA scale. Of these six, three (IRAS 16562-3959, G49.27-0.34, G339.88-1.26) display well-defined jets with knots, while the remaining two (G305A: G305.21+0.21, G305B: G305.20+0.21 and G35.58-0.03) are proposed as jet candidates based on characteristics discussed in detail in Section \ref{subsec:results}, including spectral index, morphology, and multiplicity. Additionally, source G45.12$+$0.13, has been previously reported to be associated with an UCHII region which is consistent with our assessment from Figure \ref{fig:Anglada_plot} (see below). Furthermore, this source shows ionized emission at the lowest frequency (L-band) that resembles an hour-glass morphology similar to the very massive protostar G45.47$+$0.05 which is also associated with an UC HII region that is driving a powerful photoionized outflow and a jet candidate \citep{2019ApJ...886L...4Z}.

\subsection{The Radio - Bolometric Luminosity Relation}

Here we expand on the analysis of the SOMA sample from the results presented in \citetalias{Rosero_2019}, who presented results for an initial sample of eight massive protostars. We add in the nine protostars selected from the seven high-luminosity regions analyzed in this paper. 

\begin{figure*}[ht!]
\figurenum{6}
\begin{center}
\includegraphics[width=0.488\linewidth]{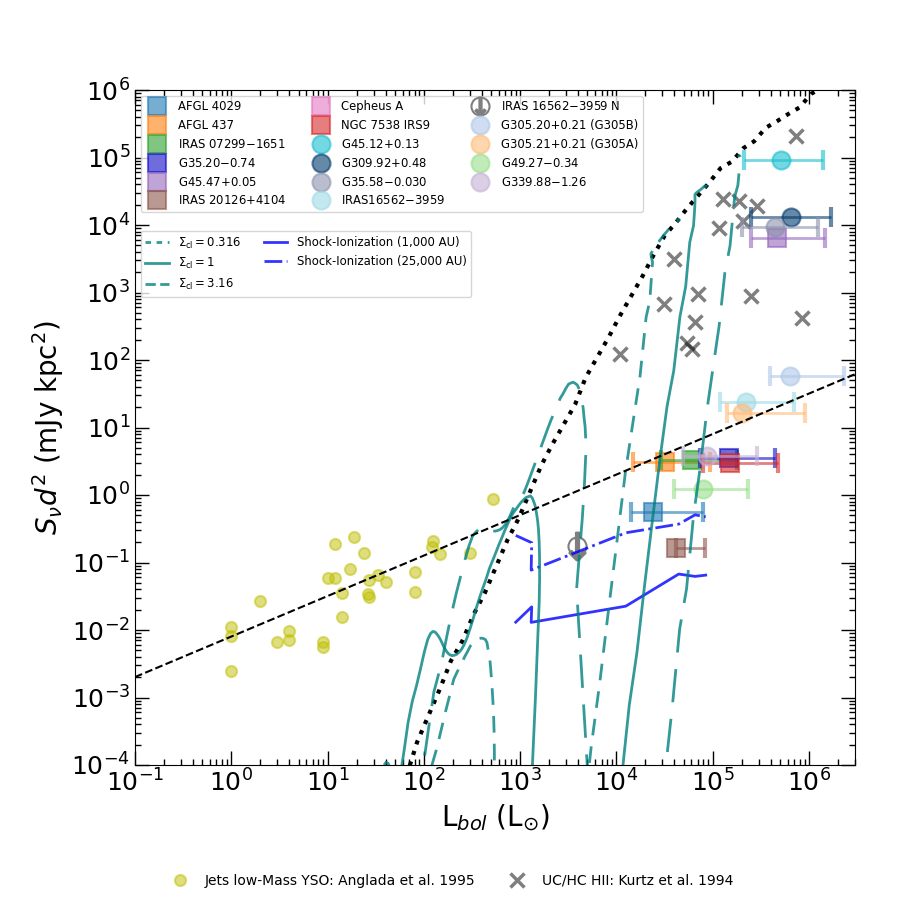} \quad \includegraphics[width=0.488\linewidth]{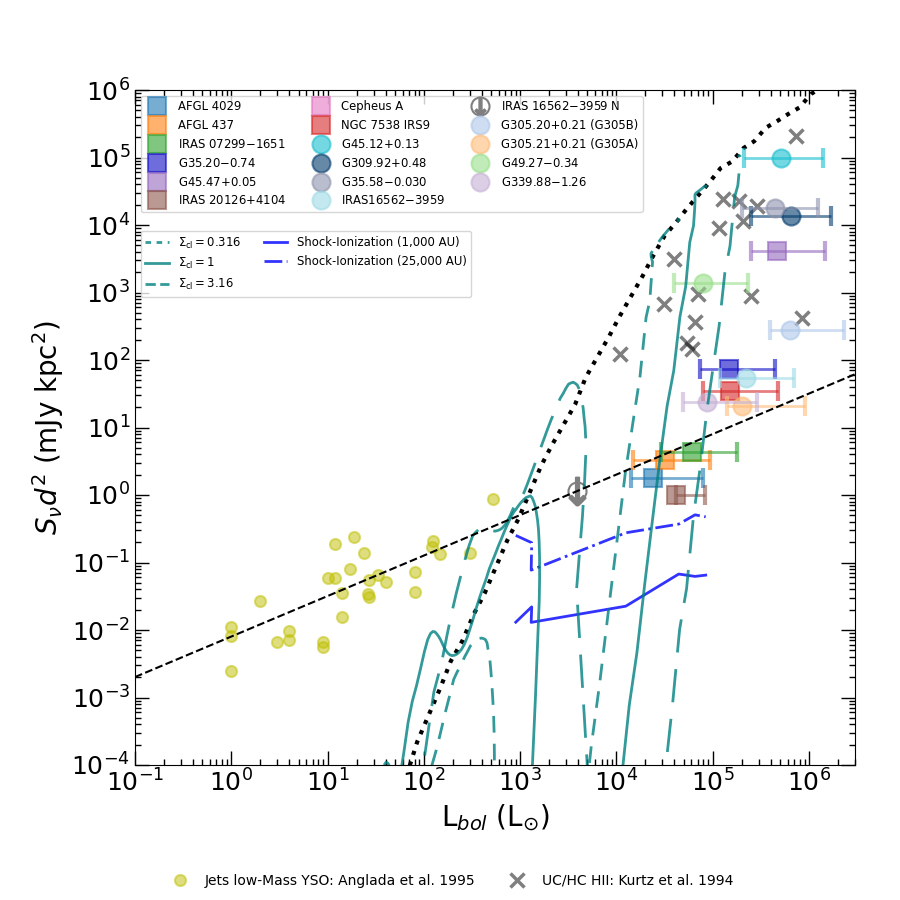}\\
\caption{Radio luminosity at 5 GHz for the Inner scale (left) and for the SOMA scale (right) as a function of the bolometric luminosity of for 15 SOMA sources from SOMA Radio I (squares) and II (circles) (see legend). In each panel we also show lower-mass YSOs from \cite{1995RMxAC...1...67A} (small yellow circles). The dashed line shows a power-law fit to these lower-mass YSOs \citep{2015aska.confE.121A}: $(S_{\nu}d^{2}/[{\rm mJy\:kpc^2}] ) = 8 \times 10^{-3}(L_{\rm bol}/L_{\odot})^{0.6}$. The $\times$ symbols are HC and UC HII regions from \cite{1994ApJS...91..659K}. The black dotted line shows the radio emission expected from optically thin HII regions powered by ZAMS stars \citep{1984ApJ...283..165T}. The light blue lines show models for HII regions powered by TCA model protostars \citep{Tanaka_2016} (with $M_{c} = 60\:M_{\odot}$ and $\Sigma_{\rm cl} =$ 0.316, 1 and 3.16~g~cm$^{-2}$, as labeled). Note that these models, which do account for radiative transfer effects of the free-free emission, assume all of the ionizing photons are reprocessed by the HII region, i.e., with zero escape fraction. The dark blue lines show radio emission from shock ionization in simulations of a fiducial TCA protostar, i.e., forming from a $60\:M_\odot$ core in a $1\:{\rm g\:cm}^{-2}$ clump environment \citep{2024ApJ...967..145G}, with the solid and dot-dashed lines showing emission from within 1,000 and 25,000~au of the protostar, respectively. \label{fig:Anglada_plot}}
\end{center}
\end{figure*}

Similar to Figure 3 in \citetalias{Rosero_2019}, in Figure \ref{fig:Anglada_plot} we present radio luminosity at 5 GHz from the Inner (left panel) and SOMA (right panel) scales versus bolometric luminosity. Note, when necessary, radio fluxes at $\nu =$ 5 GHz have been estimated by scaling from adjacent frequencies using the derived spectral index. For the bolometric luminosity of SOMA sources (detections from SOMA I and II samples are squares and circles, respectively) we report the average of the ``good'' models from Table C1 of \citetalias{Telkamp_2025} and the error bar corresponds to the dispersion of these good models. We also show data for lower-mass YSOs associated with ionized jets from \cite{1995RMxAC...1...67A} (small yellow dots). We scaled their fluxes from 3.6 cm, using a factor of 0.74, assuming that these sources have a spectral index $\alpha =$ 0.6, which is the expected value of ionized jets. A power-law fit to these data of $(S_{\nu}d^{2}/[{\rm mJy\:kpc^2}] ) = 8 \times 10^{-3}(L_{\rm bol}/L_{\odot})^{0.6}$ is shown with a dashed line. UC/HC HII regions from \cite{1994ApJS...91..659K} are represented with a $\times$symbols. Note, the bolometric luminosities of the low-mass YSOs and the UC/HC HII regions are not measured in exactly the same way as the SED-fitting method of the SOMA sources, for which we are reporting intrinsic bolometric luminosities of the fitted SED models. A plot show the results based on isotropic bolometric luminosities is presented in Appendix~\ref{sec:appA}.

We examine several theoretical models for the radio luminosity of massive protostars and show these in Figure~\ref{fig:Anglada_plot}. The black dotted line is the radio emission expected from an optically thin HII region, given the Lyman continuum luminosity of a single zero age main sequence (ZAMS) star at a given bolometric luminosity \citep{1984ApJ...283..165T}. The light blue lines corresponds to the expected radio emission that arises from photoionization from a massive protostar forming via TCA, as predicted by the \citetalias{Tanaka_2016} model, also for optically thin conditions at 5~GHz. These models have an initial core mass of $M_{c} = 60\:M_{\odot}$ and the three cases correspond to clump environment mass surface densities of $\Sigma_{\rm cl} = 0.316, 1$ and 3.16 g cm$^{-2}$. The lower mass surface densities correspond to lower accretion rates, for which protostellar contraction towards the ZAMS occurs sooner, i.e., at lower masses and lower values of $L_{\rm bol}$. Note, these models do not include any contribution from shock ionization.

A model for estimating the radio emission from free-free emission from ionized gas produced by collisional ionization in shocks has been developed for a massive protostar forming from a $60\:M_\odot$ core in a $1\:{\rm g\:cm}^{-2}$ clump environment by \citet{2024ApJ...967..145G}. This radio emission is also shown in Figure~\ref{fig:Anglada_plot} with dark blue lines. We note that the contribution from shock ionization is quite spatially extended, with the emission from a 25,000~au radius aperture (more closely matching the SOMA scale) being about a factor of 10 higher than that from a 1,000~au aperture (more closely matching the inner scale).

Comparing the distribution of SOMA Radio sources with the models, we note the following points. The SOMA Radio II sample, mostly being protostars of relatively high luminosity, help to better define the radio versus bolometric luminosity relation over the range from $L_{\rm bol}\sim 10^4$ to $10^6\:L_\odot$, which is a regime where it is rising rapidly. We find that the radio luminosity measured on the SOMA scale rises steeply by about four dex as bolometric luminosity increases by $\sim 1$ dex from $\sim10^{4.5}\:L_\odot$ to $\sim10^{5.5}\:L_\odot$. Such a steep rise is consistent with basic expectations of TCA massive protostar models.

Nevertheless, overall from the SOMA protostar sample we see that sources have relatively faint radio luminosities compared to predictions of protostellar models. This suggests that the rise in ionizing luminosity may occur only at relatively higher protostellar masses and luminosities. Such a relatively late rise in ionizing luminosity could be due to higher protostellar accretion rates or reflect systematic errors in the current protostellar models and their HII region radio emission (e.g., assumption all stellar EUV photons contribute to ionization / neglect of dust absorption). In addition, we caution that, at least on the SOMA scale, there is the possibility that we are missing radio flux due to interferometric filtering. A comparison with additional sets of theoretical models based on optically thin radio emission from the ionizing photon output of the TCA protostellar evolution models of \citet{Zhang_2018} is shown in Appendix~\ref{sec:appB}.

The differences between the results shown in the two plots of Figure~\ref{fig:Anglada_plot} (as well as Figures~\ref{fig:Anglada_plot2}, \ref{fig:YichenModels}, and \ref{fig:YichenModels2}), particularly for the sources we classified as having jet-like morphologies (G305.20+0.21, G305.21+0.21, IRAS 16562-3959, G339.88$-$1.26, and G49.27$-$0.34), arise from the different scales used to determine the radio luminosity \citep{2021A&A...645A..29K}. As noted by \cite{2024MNRAS.533.3862O}, when measuring well-resolved cores of radio jets, the sources tend to align with the fit of lower-mass YSOs associated with ionized jets (dashed line in Figure~\ref{fig:Anglada_plot}). For our inner scale (left plot), in most cases we were able to resolve the driving source of the radio jet and therefore all of our jet candidates fall in the low-mass fit within uncertainties, except for source G49.27-0.34.

The SOMA scale (right plot) takes into account all the emission in the region, adding contaminating emission from the jet knots and other sources within this region, further enhancing the radio luminosity. The artificial emission added by other sources and the different spatial scales, when compared to the inner scale, make the SOMA scale not as reliable to study the nature of the sources as the inner scale, but the SOMA scale still holds importance since it allows direct comparison with theoretical models with bigger scales that take into account jet knots and other multiple stellar sources in proto-clusters.

We note that the HC/UC HII region sample of \citet{1994ApJS...91..659K} does include sources with higher radio luminosities, potentially indicating a systematic change arising from source evolution from the protostellar phase or else reflecting biases in sample selection from a single intrinsic protostellar population that has inherent large variation in radio luminosity.

We see that the shock ionization model of \citet{2024ApJ...967..145G}, which falls below the extrapolated lower-mass YSO relation, is consistent with the fainter end of the SOMA sample. Extending the comparison of the models with SOMA data on intermediate-mass protostars is a major goal of the SOMA Radio III paper (in prep.).

\section{Summary and Conclusions}\label{summary}

We have presented results from radio continuum VLA and ATCA follow-up observations of massive protostars of the SOMA II sample following the framework developed in \citetalias{Rosero_2019}. These radio observations help to give a precise localization of the protostar, allow a search for multiplicity, and trace the presence of ionized gas, which helps trace outflow morphology, especially via orientation of radio jet axes, and is expected to be a useful evolutionary indicator of massive protostars, especially as they approach the zero age main sequence when the contribution from photoionization is predicted to increase dramatically.

A summary of main finds is as follows. We have detected and analyzed emission from 8 out of the 9 massive protostars (from 7 primary regions), with the non-detection being related to IRAS 16562-3959 N, which is an intermediate-mass protostar that is a secondary source in this region. Similar to SOMA Radio I regions, we have found a high level of ``multiplicity'' with 37 detections in the 7 regions. However, we caution that many of these sources are likely to be ionized gas jet knots, rather than independent stellar sources. Higher sensitivity observations are needed to better probe the presence of lower-mass YSOs that may be in the vicinity.

Given that the 7 primary protostars of the SOMA II sample were selected as being relatively high bolometric luminosity protostars, these data help to better define the radio - bolometric luminosity relation in this regime, i.e., from $\sim 10^4$ to $10^6\:L_\odot$. We find that the radio luminosity, i.e., at 5~GHz, rises steeply by about five dex as bolometric luminosity increases by 1 dex. However, the caveat of the possibility of missing flux from extended ionized gas features needs to be kept in mind.

Many of our fainter radio sources have radio luminosities that are consistent with a simple power law extrapolation from a lower-mass YSO sample \citep{1995RMxAC...1...67A}. However, this radio flux is about ten times higher than a theoretical model of radio emission from shock ionization of a fiducial TCA massive protostar \citep{2024ApJ...967..145G}. Nevertheless, shock ionization appears to be the most likely explanation for many of these sources, especially given the morphology of extended jet knots, while a contribution from photoionization, especially at the Inner scale, cannot be excluded. However, in the high radio luminosity regime, photoionization is expected to be dominant. Overall, these data are important for constraining theoretical models of massive protostar ionization and how this links to protostellar structure and evolutionary state.

\begin{acknowledgments}
We thank the anonymous referee for helpful comments that improved this manuscript. We acknowledge Dr. Simon Purser and Dr. Andres Guzm\'an for providing the ATCA images for  sources G339.88--1.26 and IRAS 16562--3959 presented in \citet{2016ApJ...826..208G} and \citet{2016ApJ...821...61P}, respectively. F. S. M. acknowledges support from the NRAO NINE (National and International Non-traditional Exchange) program. V. R. acknowledges support from NSF grant AST–2206437 to the Space Science Inst. J.C.T. acknowledges support from NSF grant AST–2206437 and ERC Advanced Grant 788829 (MSTAR). R.F. acknowledges support from the grants PID2023-146295NB-I00, and from the Severo Ochoa grant CEX2021-001131-S funded by MCIN/AEI/ 10.13039/501100011033 and by ``European Union NextGenerationEU/PRTR''. The National Radio Astronomy Observatory and Green Bank Observatory are facilities of the U.S. National
Science Foundation operated under cooperative agreement by Associated Universities, Inc.
\end{acknowledgments}

\facility{VLA, ATCA}

\software{CASA \citep{2007ASPC..376..127M}, Astropy \citep{2013A&A...558A..33A,2018AJ....156..123A}, APLpy \citep{aplpy2012, aplpy2019}}

\section{ORCID iDs}

\raggedright
F. Sequeira-Murillo \href{https://orcid.org/0000-0001-8169-1437}{https://orcid.org/0000-0001-8169-1437} \\
V. Rosero \href{https://orcid.org/0000-0001-8596-1756}{https://orcid.org/0000-0001-8596-1756} \\
J. Marvil \href{https://orcid.org/0000-0003-1111-8066}{https://orcid.org/0000-0003-1111-8066} \\
J. C. Tan \href{https://orcid.org/0000-0002-3389-9142}{https://orcid.org/0000-0002-3389-9142} \\
R. Fedriani \href{https://orcid.org/0000-0003-4040-4934}{https://orcid.org/0000-0003-4040-4934} \\
Y. Zhang \href{https://orcid.org/0000-0001-7511-0034}{https://orcid.org/0000-0001-7511-0034} \\
A. Robinson \href{https://orcid.org/0009-0007-4080-9807}{https://orcid.org/0009-0007-4080-9807} \\
P. Gorai \href{https://orcid.org/0000-0003-1602-6849}{https://orcid.org/0000-0003-1602-6849} \\
K. E. I. Tanaka \href{https://orcid.org/0000-0002-6907-0926}{https://orcid.org/0000-0002-6907-0926} \\
M. Liu \href{https://orcid.org/0000-0001-6159-2394}{https://orcid.org/0000-0001-6159-2394} \\
J. M. De Buizer \href{https://orcid.org/0000-0001-7378-4430}{https://orcid.org/0000-0001-7378-4430} \\
M. T. Beltran \href{https://orcid.org/0000-0003-3315-5626}{https://orcid.org/0000-0003-3315-5626} \\
R. D. Boyden \href{https://orcid.org/0000-0001-9857-1853}{https://orcid.org/0000-0001-9857-1853} \\

\appendix

\section{Radio luminosity versus isotropic bolometric luminosity}\label{sec:appA}

\begin{figure*}[ht!]
\figurenum{A1}
\begin{center}
\includegraphics[width=0.488\linewidth]{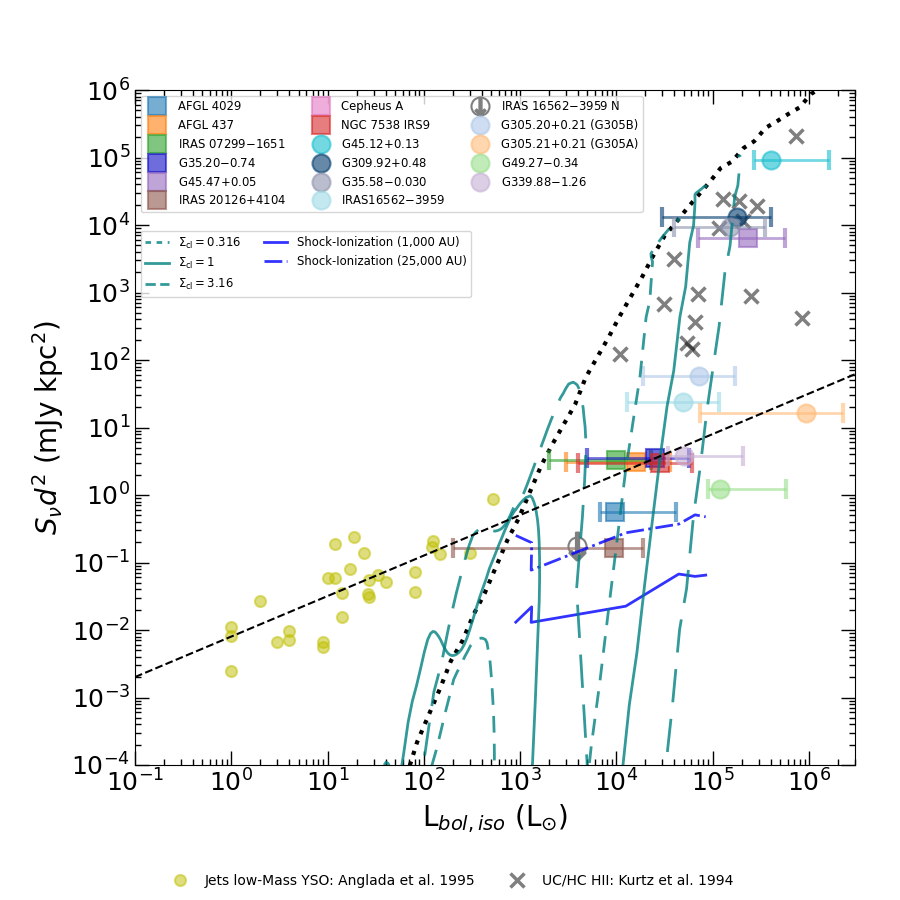}\quad \includegraphics[width=0.488\linewidth]{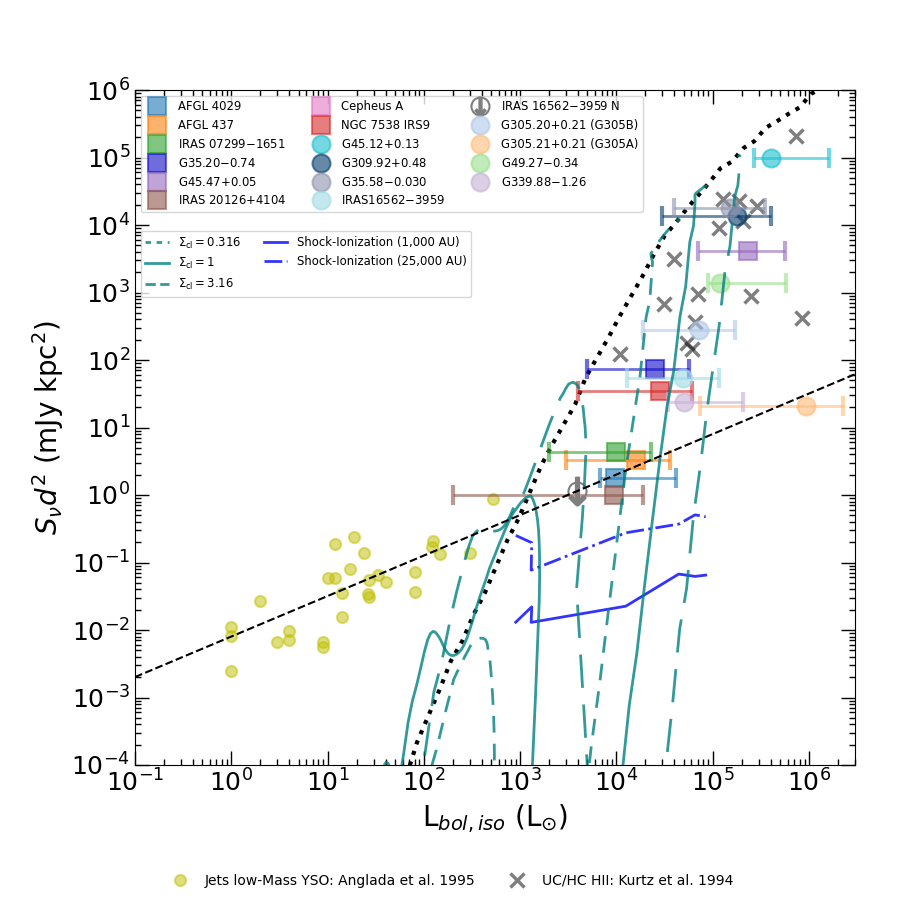}
\caption{As Fig.~\ref{fig:Anglada_plot}, but now plotting $L_{\rm bol,iso}$ for the SOMA sources (left: inner scale, right: SOMA scale).
\label{fig:Anglada_plot2}}
\end{center}
\end{figure*}

Figure \ref{fig:Anglada_plot2} shows a similar plot to Figure \ref{fig:Anglada_plot}, but in this case we used the average of the good models of the isotropic bolometric luminosity (Table C.1 of \citetalias{Telkamp_2025}). Everything else is the same as Figure \ref{fig:Anglada_plot}.

\section{Evolutionary tracks from radio luminosities}\label{sec:appB}

\begin{figure*}[ht!]
\figurenum{B1}
\begin{center}
\includegraphics[width=0.32\linewidth]{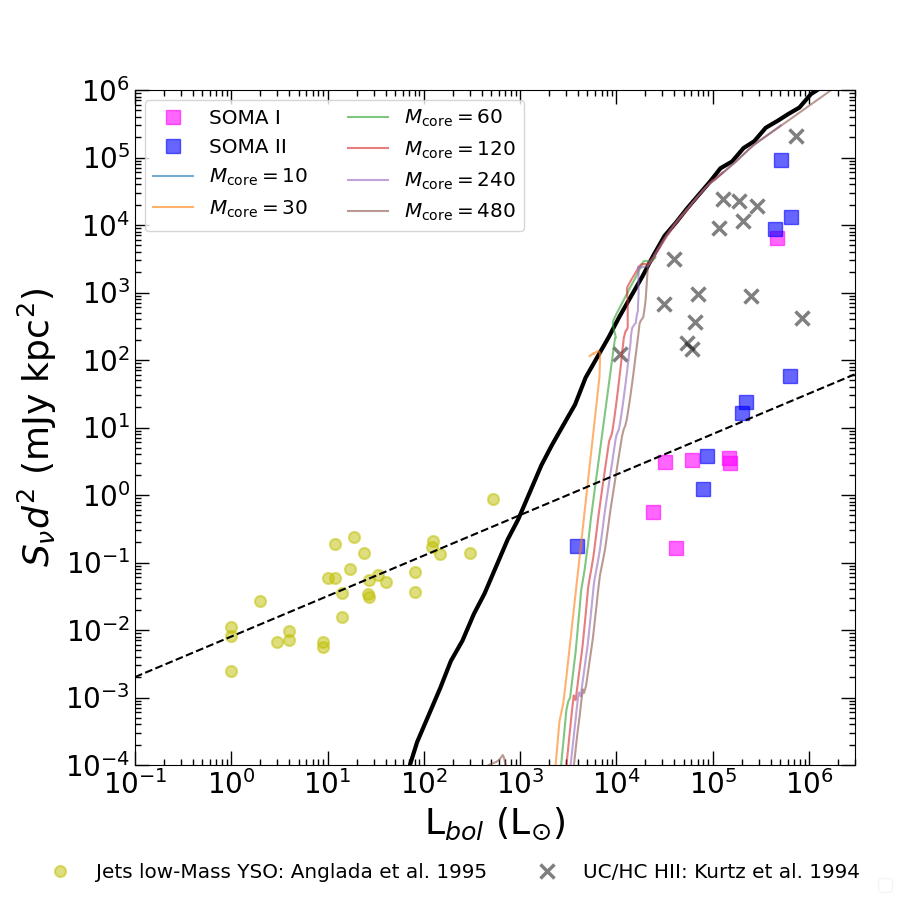}\quad\includegraphics[width=0.32\linewidth]{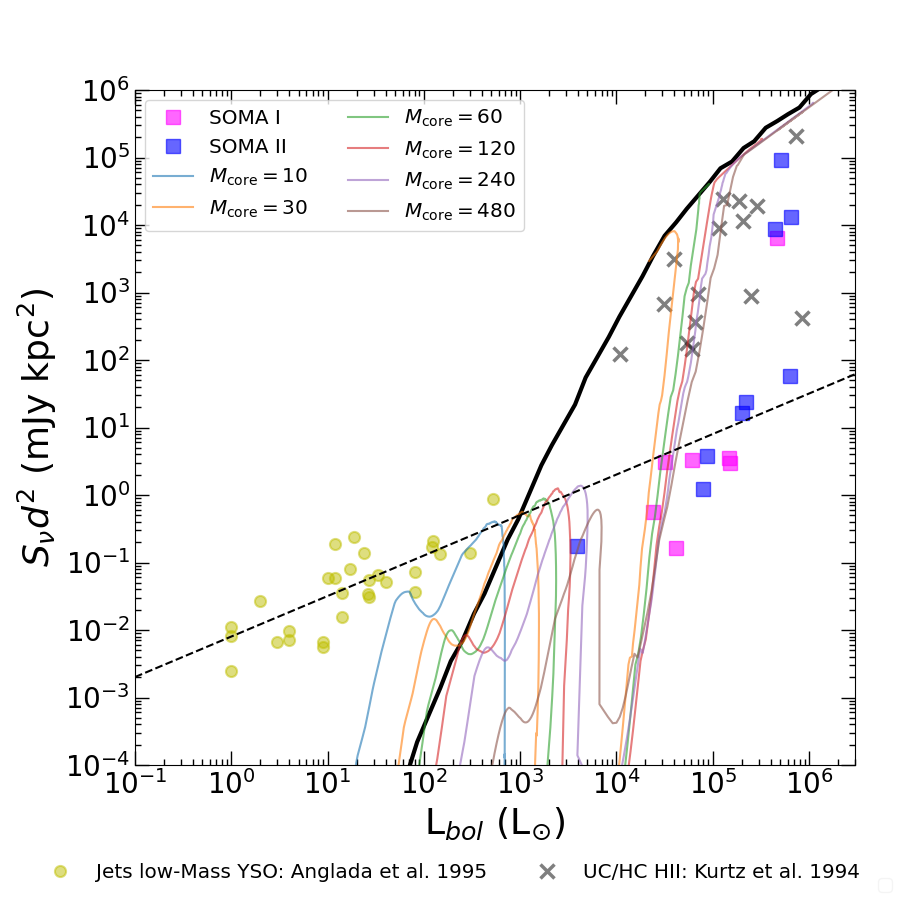}\quad\includegraphics[width=0.32\linewidth]{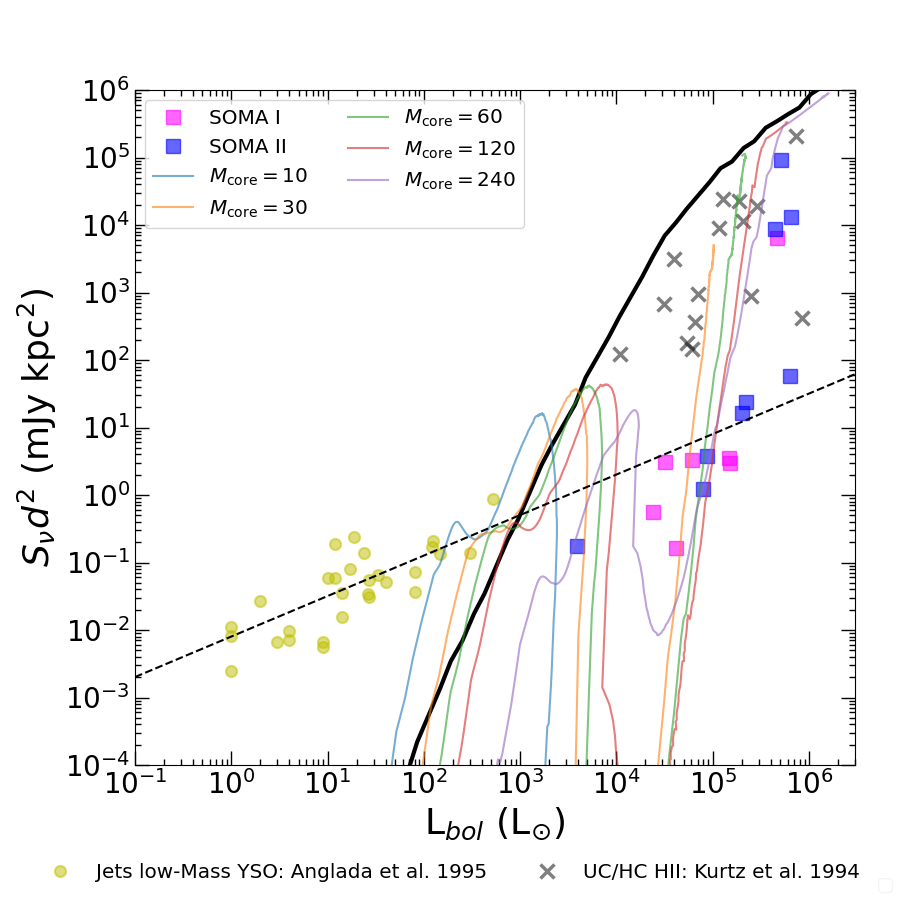}\\
\includegraphics[width=0.32\linewidth]{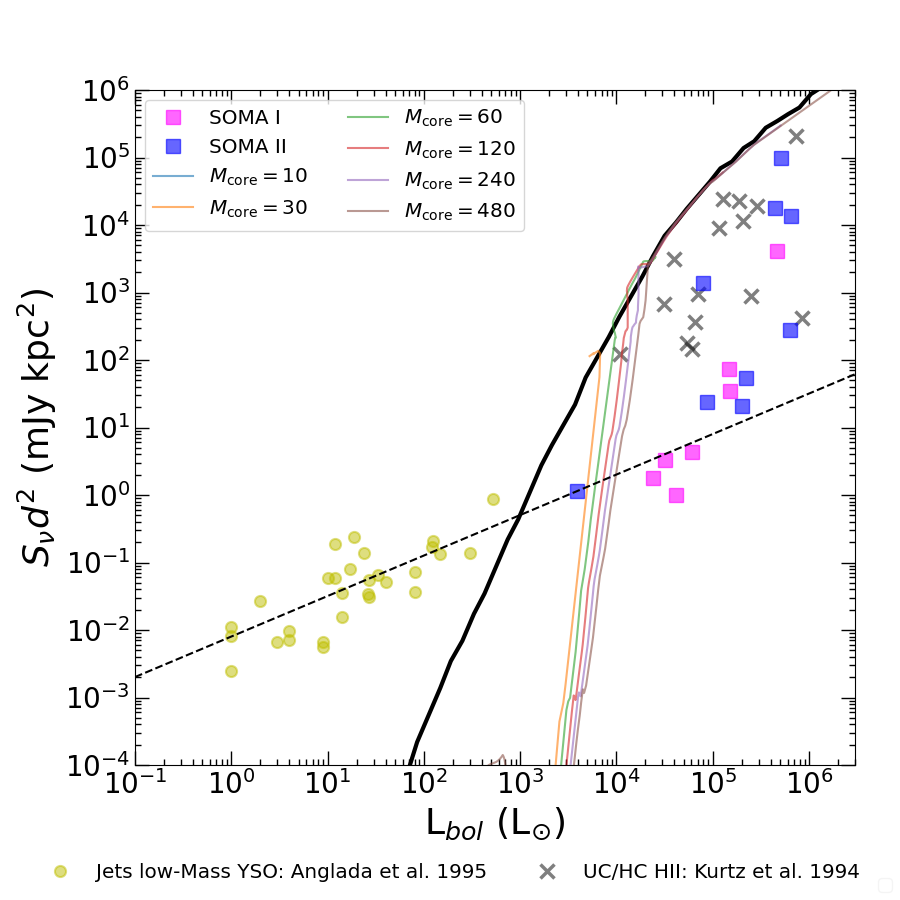}\quad\includegraphics[width=0.32\linewidth]{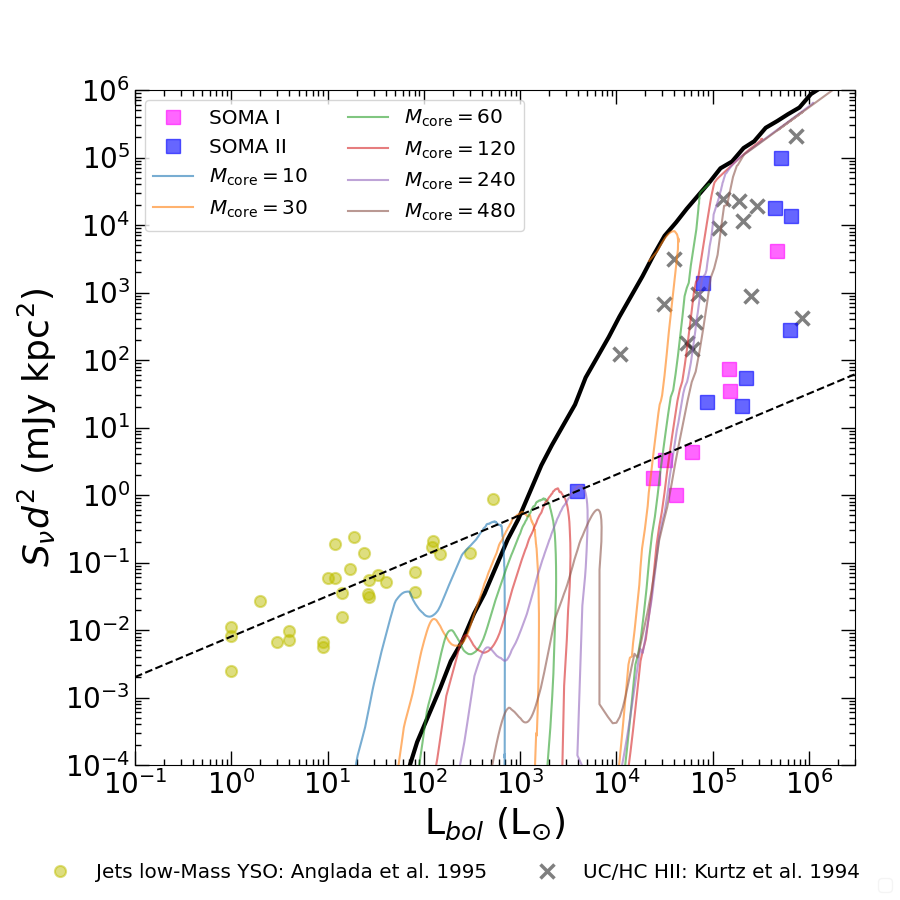}\quad\includegraphics[width=0.32\linewidth]{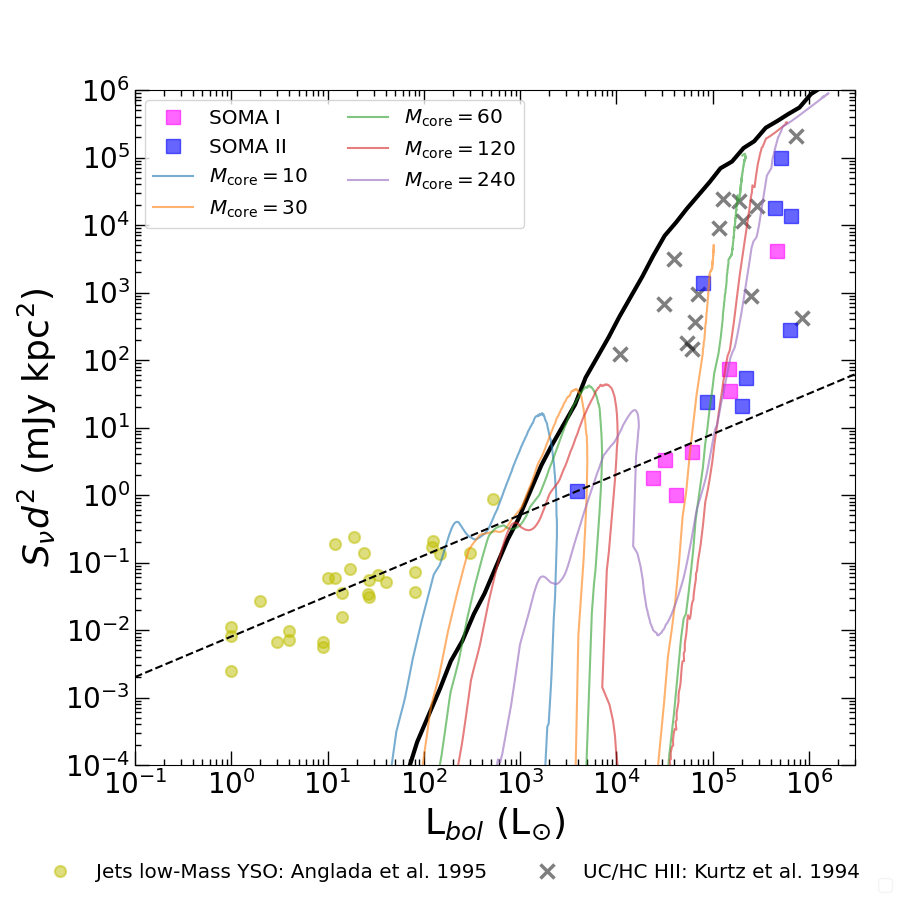}
\caption{As Fig.~\ref{fig:Anglada_plot}, but now showing protostellar evolutionary tracks from \citet{Zhang_2018}, under the assumption of optically thin radio emission. These models are for $M_{c} =$ 10, 30, 60, 120, 240 and 480 $M_{\odot}$ with $\Sigma_{cl} =$ 0.316 g cm$^{-2}$ (left), 1 g cm$^{-2}$ (center) and 3.16 g cm$^{-2}$ (right). Top plots are with the radio luminosity at the inner scale and bottom plots are with the radio luminosity at the SOMA scale.
\label{fig:YichenModels}}
\end{center}
\end{figure*}

\begin{figure*}[ht!]
\figurenum{B2}
\begin{center}
\includegraphics[width=0.32\linewidth]{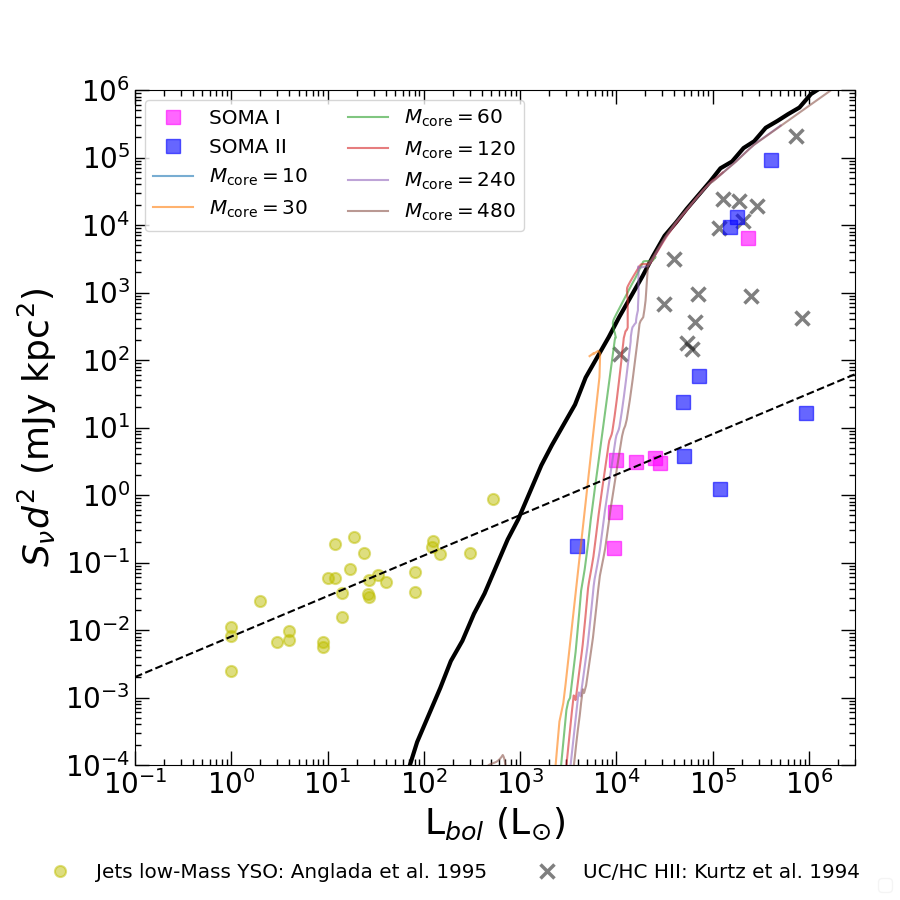}\quad\includegraphics[width=0.32\linewidth]{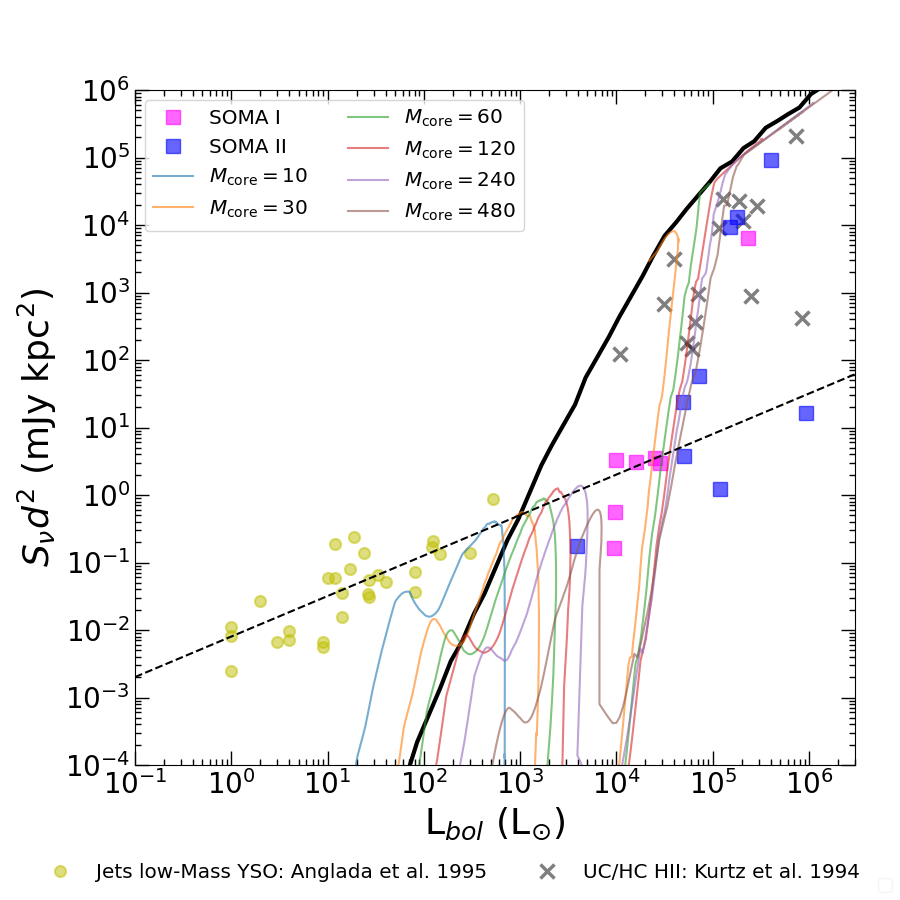}\quad\includegraphics[width=0.32\linewidth]{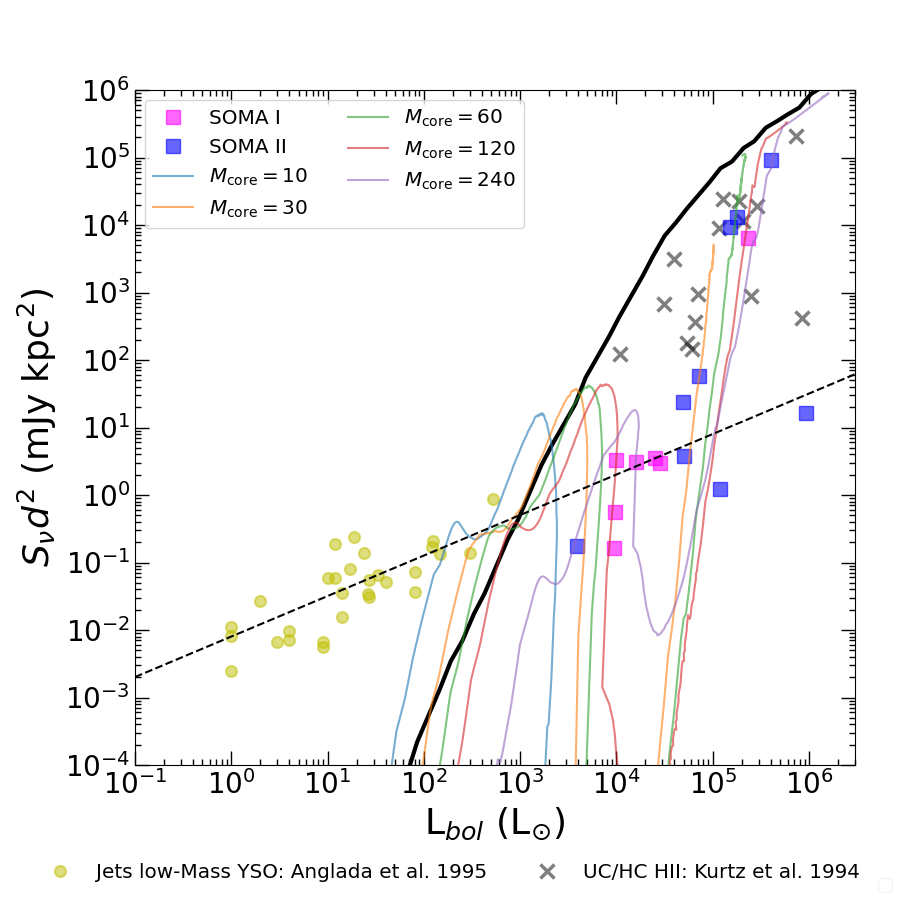}\\
\includegraphics[width=0.32\linewidth]{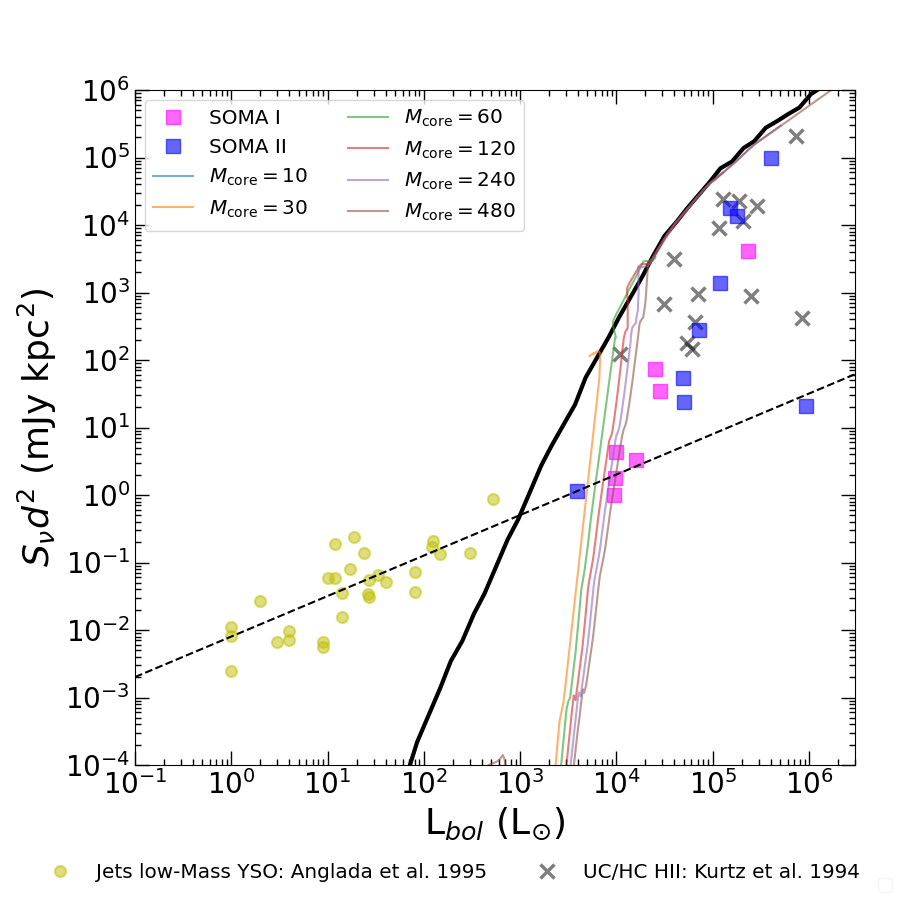}\quad\includegraphics[width=0.32\linewidth]{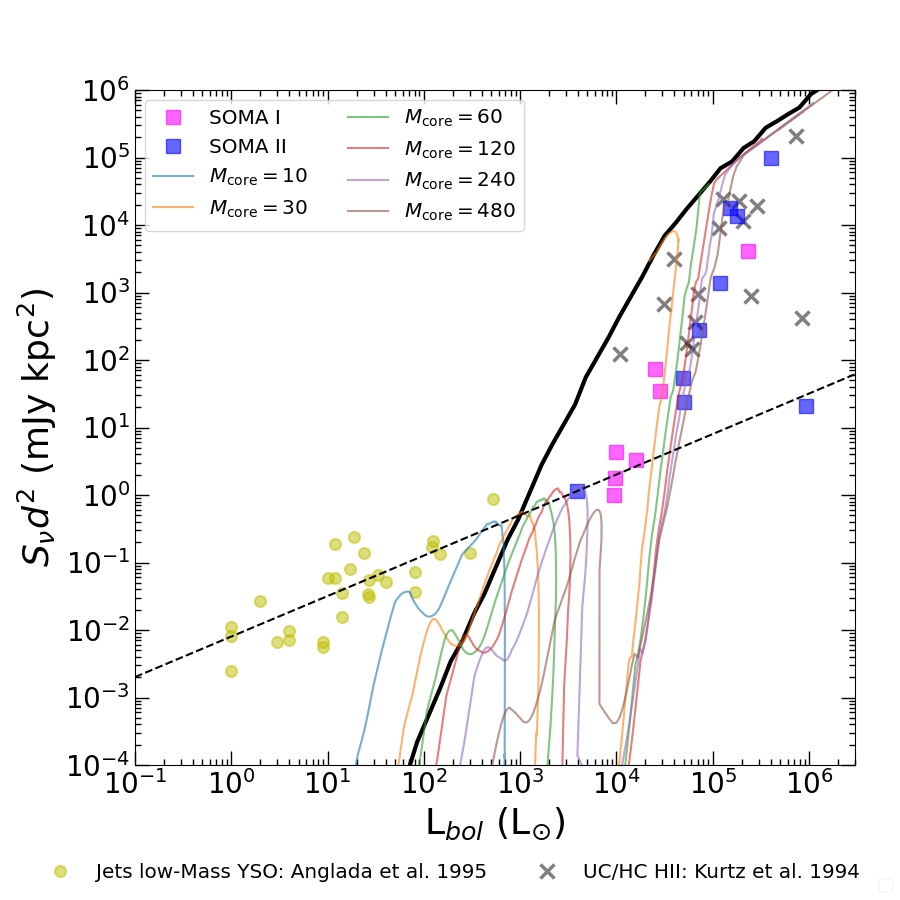}\quad\includegraphics[width=0.32\linewidth]{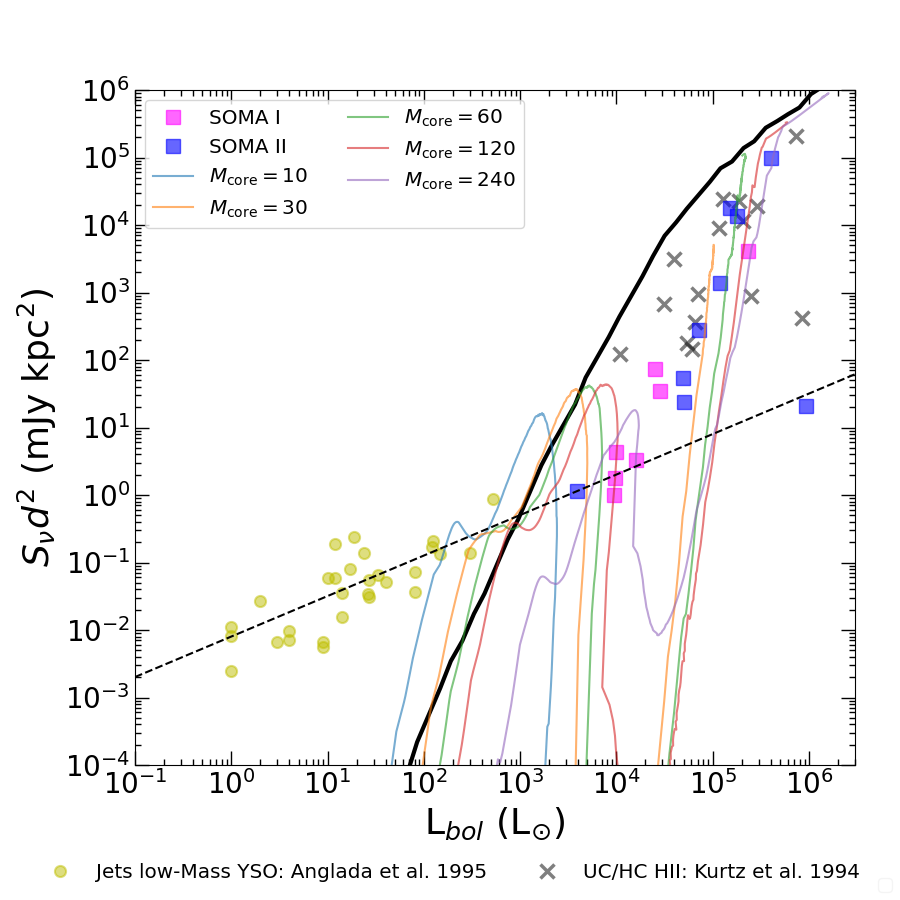}
\caption{As Fig.~\ref{fig:YichenModels}, but now with the isotropic bolometric luminosity. Top plots are with the radio luminosity at the inner scale and bottom plots are with the radio luminosity at the SOMA scale.
\label{fig:YichenModels2}}
\end{center}
\end{figure*}

Figure \ref{fig:YichenModels} makes a comparison of the SOMA Radio data at the inner (top) and SOMA (bottom) scale, with the full set of protostellar evolutionary tracks from \citet{Zhang_2018}, i.e., with $M_{c} =$ 10, 30, 60, 120, 240 and 480 $M_{\odot}$ with $\Sigma_{cl} =$ 0.316 g cm$^{-2}$ (left), 1 g cm$^{-2}$ (center) and 3.16 g cm$^{-2}$ (right). In these models the radio luminosities are not set from radiation transfer like the tracks from \citep{Tanaka_2016}, but instead assumes a simple spherical HII region. Meanwhile, Figure \ref{fig:YichenModels2} shows the same information but with the average of the good models of the isotropic bolometric luminosity (Table C.1 of \citetalias{Telkamp_2025}).

\newpage
\bibliography{sample701}{}
\bibliographystyle{aasjournalv7}

\end{document}